\documentclass[]{pasj01}
\usepackage{comment}
\usepackage[switch,mathlines]{lineno}
\usepackage{here}
\usepackage{array}
\usepackage{amsthm}
\usepackage{tabularx}
\usepackage{multirow}
\Received{$\langle$reception date$\rangle$}
\Accepted{$\langle$acception date$\rangle$}
\Published{$\langle$publication date$\rangle$}

\begin{document}

\title{Components of star formation in NGC\,253 : Non-negative Matrix Factorization Analysis with the ALCHEMI integrated intensity images}
\author{Ryo \textsc{Kishikawa},\altaffilmark{1}$^{*}$
        Nanase \textsc{Harada},\altaffilmark{2,3}
        Toshiki \textsc{Saito},\altaffilmark{4}
        Susanne \textsc{Aalto},\altaffilmark{5}
        Laura \textsc{Colzi},\altaffilmark{6}
        Mark \textsc{Gorski},\altaffilmark{7}
        Christian \textsc{Henkel},\altaffilmark{8,9}
        Jeffrey G.  \textsc{Mangum},\altaffilmark{10}
        Sergio \textsc{Mart\'in},\altaffilmark{11,12}
        %David S. \textsc{Meier},\altaffilmark{13,14}
        Sebastian \textsc{Muller},\altaffilmark{5}
        Yuri \textsc{Nishimura},\altaffilmark{13}
        V\'ictor  M. \textsc{Rivilla},\altaffilmark{6}
        Kazushi \textsc{Sakamoto},\altaffilmark{14}
        Paul \textsc{van der Werf},\altaffilmark{15}
        Serena \textsc{Viti}, \altaffilmark{15,16}
        }
\altaffiltext{1}{Department of Physics, Tokyo Metropolitan University, 1-1 Minami-Osawa, Hachioji, Tokyo 192-0397, Japan}
\altaffiltext{2}{National Astronomical Observatory of Japan, 2-21-1 Osawa, Mitaka, Tokyo 181-8588, Japan}
\altaffiltext{3}{Astronomical Science Program, Graduate Institute for Advanced Studies, SOKENDAI, 2-21-1 Osawa, Mitaka, Tokyo 181-1855, Japan}
\altaffiltext{4}{Faculty of Global Interdisciplinary Science and Innovation, Shizuoka University, 836 Ohya, Suruga-ku, Shizuoka 422-8529, Japan}
\email{kishikawa-ryo@ed.tmu.ac.jp}
\altaffiltext{5}{Department of Space, Earth and Environment, Chalmers University of Technology, Onsala Space Observatory, SE-439 92 Onsala, Sweden}
\altaffiltext{6}{Centro de Astrobiología (CAB), CSIC-INTA, Ctra. de Ajalvir Km. 4, 28850, Torrejón de Ardoz, Madrid, Spain}
\altaffiltext{7}{Center for Interdisciplinary Exploration and Research in Astrophysics (CIERA) and Department of Physics and Astronomy, Northwestern University, Evanston, IL 60208, USA}
\altaffiltext{8}{Max-Planck-Institut f\"ur Radioastronomie, Auf dem H\"ugel 69, 53121 Bonn, Germany}
\altaffiltext{9}{Xinjiang Astronomical Observatory, Chinese Academy of Sciences, 830011, Urumqi, PR China}
\altaffiltext{10}{National Radio Astronomy Observatory, 520 Edgemont Road, Charlottesville, VA  22903-2475, USA}
\altaffiltext{11}{European Southern Observatory, Alonso de C\'ordova, 3107, Vitacura, Santiago 763-0355, Chile}
\altaffiltext{12}{Joint ALMA Observatory, Alonso de C\'ordova, 3107, Vitacura, Santiago 763-0355, Chile}
%\altaffiltext{13}{New Mexico Institute of Mining and Technology, 801 Leroy Place, Socorro, NM 87801, USA}
%\altaffiltext{14}{National Radio Astronomy Observatory, PO Box O, 1003 Lopezville Road, Socorro, NM 87801, USA}
\altaffiltext{13}{Institute of Astronomy, Graduate School of Science,
The University of Toky, 2-21-1 Osawa, Mitaka, Tokyo 181-0015, Japan}
\altaffiltext{14}{Institute of Astronomy and Astrophysics, Academia Sinica, 11F of AS/NTU
Astronomy-Mathematics Building, No.1, Sec. 4, Roosevelt Rd, Taipei 106216, Taiwan}
\altaffiltext{15}{Leiden Observatory, Leiden University, P.O. Box 9513, 2300 RA Leiden, The Netherlands}
\altaffiltext{16}{Physics and Astronomy Department, University College London, Gower St, London, WC1E 6BT, UK}

\KeyWords{galaxies: starburst --- galaxies: star formation --- galaxies: individual (NGC 253) --- ISM: molecules --- methods: data analysis}

\maketitle

\begin{abstract}
It is essential to examine the physical or chemical properties of molecular gas in starburst galaxies to reveal the underlying mechanisms characterizing starbursts. We used non-negative matrix factorization (NMF) to extract individual molecular or physical components involved in the star formation process in NGC\,253. We used images of 148 transitions from 44 different species of the ALMA large program ALCHEMI. Additionally, we included the continuum images at ALMA Bands 3 and 7 from the same dataset. For the five NMF components (NF1--5), we obtained that their distributions correspond to various basic phenomena related to star formation: i) low-density gas extended through the galactic central molecular zone (NF2), ii) shocks (NF3), iii) starburst regions (NF4), and iv) young star-forming regions (NF5). The other component (NF1) is related to excitation; three components obtained by NMF (NF3, 1, and 5) show a strong dependence upon the upper state energies of transitions, and represent low-, intermediate-, and high-excitation, respectively. We also compared our results using principal component analysis (PCA) previously applied to the same dataset. Molecular components extracted from NMF are similar to the ones obtained from PCA. However, NMF is better at extracting components associated with a single physical component, while a single component in PCA usually contains information on multiple physical components. This is especially true for features with weak intensities like emission from outflows. Our results suggest that NMF can be one of promising methods interpreting molecular line survey data, especially in the upcoming era of wide-band receivers.
\end{abstract}
% \pagewiselinenumbers

\section{Introduction}
Star formation in galaxies is one of the principal elements of galaxy formation and evolution. In the past Universe, around redshifts of $z\sim 2-3$, star formation was more active than now (e.g., \cite{Madau2014}). Although the current Universe hosts a lower average star formation activity than that during these active cosmic periods, galaxies hosting events of elevated star formation activity still exist (i.e., starburst galaxies). These local starburst galaxies allow for glimpses of the past universe when star formation was more active (e.g., \cite{Weedman1981}; \cite{Kennicutt2012}). The amount of gas is one of the factors that influences the star formation rate (SFR). Specifically, it is the gas in the molecular form that is dense and cold enough to allow for star formation. It is well known that stellar bars or galaxy mergers can transfer significant amounts of gas into the galaxy center (e.g., \cite{Sakamoto1999} and \cite{Saito2015}). Other factors that affect the SFR are the physical properties of the molecular gas. For example, stars preferentially form in dense gas (e.g., \cite{Gao2004} and \cite{Lada2010}). Starbursts are suppressed by large amounts of ultraviolet stellar radiation and/or by outflows associated with star formation. This process is called negative feedback \citep{Grudic2018}, and is considered important in quenching star formation. While the molecular gas mass can be measured with low-$J$ CO transitions, it is difficult to measure the physical properties of the different molecular components involved in the process (e.g., shocks, UV radiation, cosmic-ray ionization, etc.).

Astrochemistry can be used to probe these physical properties. As stars form by the contraction of molecular clouds, the nature of such molecular gas and the effects of feedback can be investigated through observations of various molecular species from infrared to millimeter waves \citep{Yamamoto2017}. Although it has been difficult to image molecular transitions weaker than a few bright species (e.g., CO, HCN, HCO$^+$, etc.) in external galaxies, the Atacama Large Millimeter/submillimeter Array (ALMA) has enabled high angular resolution and sensitive observations of a larger variety of species \citep{Meier2015}. The large program ALMA Comprehensive High-resolution Extragalactic Molecular Inventory (ALCHEMI) survey has detected over 1500 molecular and radio recombination lines (RRLs) in NGC\,253 \citep{Martin2021}. NGC\,253 is a starburst galaxy with a SFR of $\sim 2 \> \MO \> \mathrm{yr^{-1}}$ in the central region \citep{Leroy2015}. It is located at a distance of $3.5 \> \mathrm{Mpc}$ \citep{Rekola2005}, one of the closest starburst galaxies to our Galaxy that is not strongly affected by an active galactic nucleus (AGN) (\cite{Brunthaler2009}; \cite{Muller2010}). There are some giant molecular clouds (GMCs) in the central region (e.g., \cite{Sakamoto2011}; \cite{Leroy2015}). \citet{Ando2017} revealed that eight star-forming clumps in the center of the galaxy are chemically distinct despite the similarity in mass and size. Evolutionary stages of star formation and environmental differences of molecular clouds may be apparent in the chemical composition. By examining molecular species and energy levels of their transitions in NGC\,253, we may gain new insights into our understanding of starburst phenomena. In previous research, ALCHEMI has revealed the mechanism and properties of shocks in NGC\,253 (\cite{Huang2023}; \cite{Harada2022}; \cite{Humire2022}). The survey also found that the influence of cosmic rays is high using specific molecular species (\cite{Holdship2021}; \cite{Harada2021}; \cite{Holdship2022}; \cite{Behrens2022}). Some of the rarest species, such as PN, were detected for the first time outside of the Galaxy (\cite{Haasler2022}). Other studies have confirmed the physical properties of an outflow (\cite{Bao2024}), and the difference in density structure between the central molecular zones of NGC\,253 and that of our own Galaxy (\cite{Tanaka2024}). \citet{2024arXiv240508408B} investigated abundances of sulfur-bearing species to measure how much sulfur is in the gas phase, which is largely unknown. \citet{2024A&A...686A..31B} compared the cluster age with the isotopic ratio $^{12}$C/$^{13}$C from isotopologues of CO, HCN, and HCO$^+$. The complex dataset also allowed us to train a neural network to probe the automatic molecular parameter determination \citep{Barrientos2021}.

The amount of information contained in such molecular line surveys can be overwhelming. Machine learning is a powerful method when we analyze such big data. Among them, dimensionality reduction, one of the unsupervised learning methods, is an effective way to extract features from a large amount of data. Dimensionality reduction has various methods such as principal component analysis (PCA), non-negative matrix factorization (NMF), and t-distributed stochastic neighbor embedding (t-SNE). The astronomical community has already leveraged these methods \citep{1984MNRAS.206..453E,2014MNRAS.437..241H,2017ApJS..228...24T} for the interpretation of the data. PCA has also been used to reduce the dimensionality of integrated intensity images of molecular transitions by many studies (e.g., \cite{Meier2005}, \cite{Watanabe2016}, \cite{Gratier2017}, \cite{Saito2022}, \cite{Harada2024}). Using the PCA method to an unprecedented scale, \citet{Harada2024} succeeded in extracting some components involved in star formation by applying PCA to pixel data of $\sim 150$ integrated intensity (moment 0) maps of multi-transition lines in NGC\,253 using the ALCHEMI data. However, multiple physical components could appear in a single principal component because principal components are orthogonal to each other while physical components may affect intensities in a non-orthogonal way.
%some of the components have become mixed and interpretation difficulties exist due to mixed positive and negative values of components. 

Recently, \citet{Mijolla2024} applied NMF on simulated data, and demonstrated its ability to distinguish molecular emission coming from different density components. Although most studies of molecular transitions used PCA for dimensionality reduction, here we explore the NMF method to extract components uniquely related to molecular/physical components. In this study, we apply NMF to the ALCHEMI data. NMF is one of the strong dimensionality reduction methods for data without negative values (\cite{Lee1999}; \cite{Lee2001}). It is likely effective for astronomical images because their signals usually have non-negative values, unless there are absorption lines in the data. In molecular line images, there are multiple components with varying physical parameters such as temperature, density, total hydrogen column densities, and molecular compositions. An addition of a cloud with certain physical parameters should only increase the emission intensity, and not decrease it if we can ignore absorption. Therefore, components obtained by NMF are expected to represent more unique components representing certain physical parameters than PCA. 
%Given a single non-negative matrix, NMF finds two matrices with non-negative elements numerically. we tried to extract the structure associated with star formation by using emission lines using NMF. We expect it to be easy to interpret components after applying NMF because this method does not handle negative values, unlike other methods. 
%A previous work on NMF applied to simulated molecular emission data showed that this method was effective in extracting various density components \citep{Mijolla2024}. 
% Knowing some of the structure, such as collisions between molecular clouds and feedbacks associated with star formation, can give us an insight into current and future star formation. If we can extract the structure that gives rise to star formation and find some emission lines that explain the structure well, it will lead to further understanding of star formation in starburst galaxies. 

This paper is structured as follows. In Section \ref{sec:method}, we describe the data and how we analyze them using NMF. In Section \ref{sec:results}, we show the distribution and dependence of our results on the energy level of transition after applying NMF. In Section \ref{sec:discussion}, we discuss the interpretation of each component and tracers involved in star formation. Further, we discuss the advantages of NMF over PCA. We end with a summary of the main results of the analysis in Section \ref{sec:summary}.

\begin{figure*}[ht]
    \begin{center}
    \includegraphics[keepaspectratio, width=\linewidth]{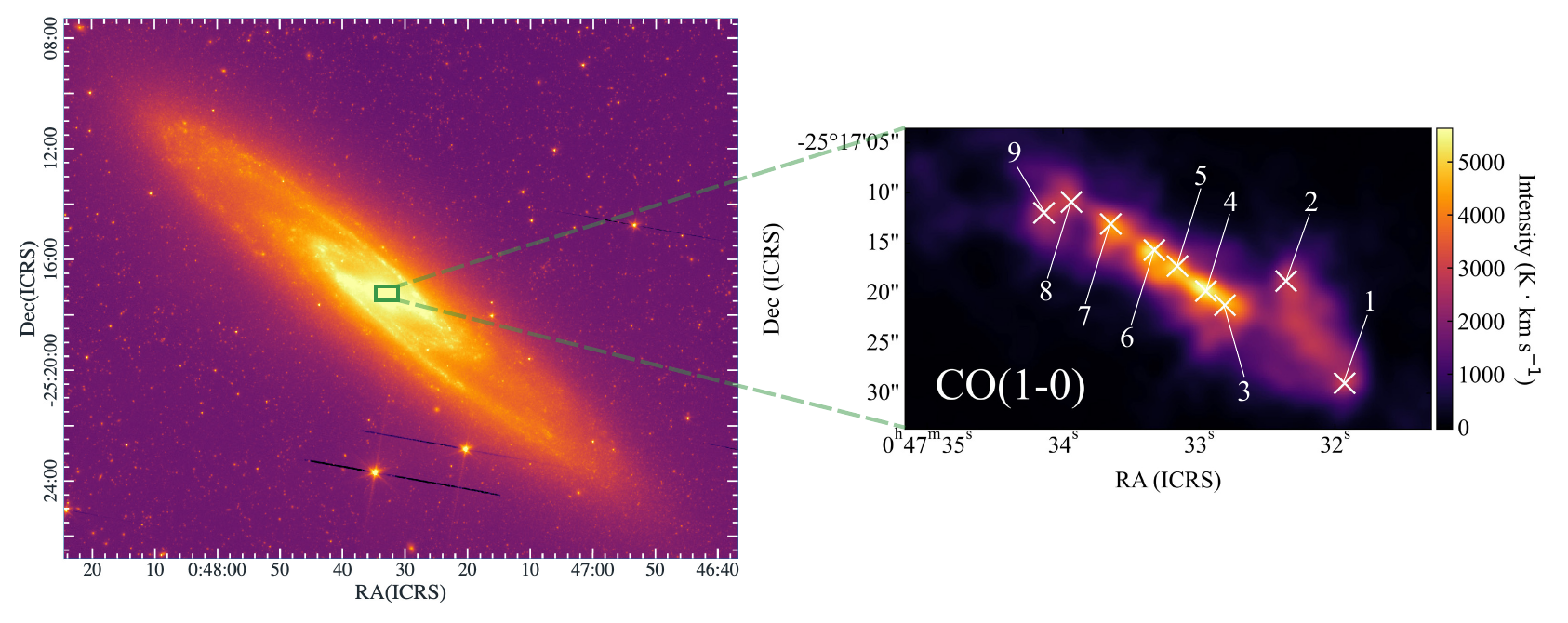}
    \end{center}
\caption{The image on the left is captured by the Spitzer Space Telescope (SST). The image on the right is a velocity-integrated map of CO(1--0) by the ALCHEMI survey. Nine GMCs with different chemical properties exist in this region (See table \ref{table:GMCs}).}
\label{fig:region}
\end{figure*}

\begin{table*}[ht]
    \tbl{Each GMC position referred in this study (taken from \cite{Harada2024}).}{%}
    \centering
    \begin{tabular}{cccc}
    \hline
    \begin{tabular}{c} GMC \\ number \end{tabular} & \begin{tabular}[c]{@{}c@{}}RA(ICRS) \\($00^{\mathrm{h}}:47^{\mathrm{m}}:-^{\mathrm{s}}$)\end{tabular} & \begin{tabular}[c]{@{}c@{}}Dec(ICRS) \\($-25^{\mathrm{\circ}}:17^{\mathrm{'}}:-^{\mathrm{''}}$)\end{tabular} & Remark \\ \hline
    1 & $31.93$ & $29.0$ & Class I methanol maser\footnotemark[$*$] \\
    2 & $32.36$ & $18.8$ & Class I methanol maser\footnotemark[$*$] \\
    3 & $32.81$ & $21.2$ & Clumps 1-3\footnotemark[$\dag$] \\
    4 & $32.95$ & $19.8$ & Clumps 4-7\footnotemark[$\dag$]\\
    5 & $33.16$ & $17.3$ & Clumps 8-13\footnotemark[$\dag$], the TH2 region\footnotemark[$\ddag$]\\
    6 & $33.33$ & $15.7$ & Clump 14\footnotemark[$\dag$]\\
    7 & $33.65$ & $13.1$ & Class I methanol maser\footnotemark[$*$] \\
    8 & $33.94$ & $08.9$ & Class I methanol maser\footnotemark[$*$] \\
    9 & $34.14$ & $12.0$ & Class I methanol maser\footnotemark[$*$] \\ \hline
    \end{tabular}}
    \begin{tabnote}
        \footnotemark[$*$] \citet{Humire2022}
        \footnotemark[$\dag$] \citet{Leroy2018}
        \footnotemark[$\ddag$]
        \citet{Turner1985}
    \end{tabnote}
    \label{table:GMCs}
\end{table*}

\section{Data and Method}\label{sec:method}
\begin{figure*}[ht]
    \begin{center}
    \includegraphics[keepaspectratio, width=\linewidth]{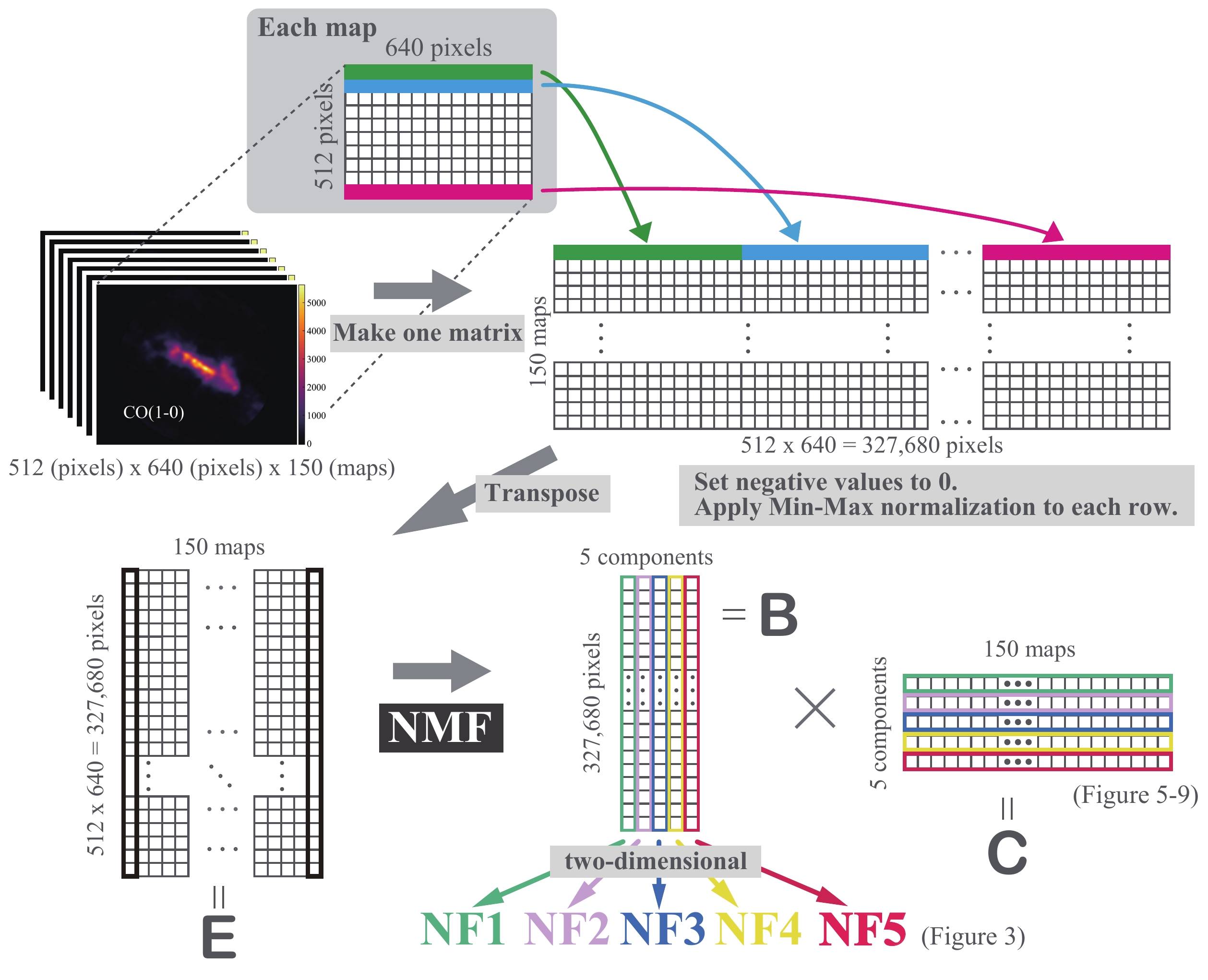}
    \end{center}
\caption{Schematic representation of the methodology of this study.}
%Integrated-velocity maps (CO (1-0) and 0.8\,mm continuum as example) are zoomed to include the central molecular zone.
\label{method}
\end{figure*}
\subsection{ALCHEMI data}
The ALCHEMI survey (\cite{Martin2021}) observed the central molecular zone (CMZ) of NGC\,253 with the field of view of $50^{"} \times 20^{"} (= 850 \mathrm{\>pc} \times 340 \mathrm{\>pc}$) with an angular resolution of $1.6^{"}$ (= 27\,pc). The survey consisted of a spectral scan over most of the frequency coverage in ALMA Bands 3-7. We took exactly the same integrated intensity images used in the PCA study by \citet{Harada2024}, which omitted transitions with blending above certain levels. We used two-dimensional data ($512 \mathrm{\>pixels}\times 640 \mathrm{\>pixels}$) of 150 images (molecular lines, recombination lines, and continuum emission) with a pixel size of $0\farcs15$ as schematically shown in figure \ref{method}. An complete list of transitions can be found in Table 3 of \citet{Harada2024}. A integrated intensity image and the region taken by the ALCHEMI survey are shown in figure \ref{fig:region}. The positions of the GMCs of NGC~253 are provided in table \ref{table:GMCs}.

\subsection{NMF}\label{subsec: NMF}

We explain here the flow of dimensionality reduction using non-negative matrix factorization (NMF). The basic concept of NMF is to factorize an input matrix \textsf{E} into two non-negative matrices \textsf{B} and \textsf{C} to satisfy the condition
\begin{eqnarray}
    \bm{\textsf{E}} \simeq \bm{\textsf{B}} \bm{\textsf{C}},
\end{eqnarray}

where \textsf{E} is an $L\times N$ matrix, \textsf{B} is an $L\times M$ matrix, and \textsf{C} is an $M\times N$ matrix. With this factorization using $M<\mathrm{min}(L,N)$, the dimension of \textsf{E} is reduced from $N$ to $M$. Note that it is usually not possible to find two matrices $\bm{\textsf{B}}$ and $\bm{\textsf{C}}$ whose multiplication exactly equals to $\bm{\textsf{E}}$. We instead minimize the error between them. The exact method of minimization is found in appendix \ref{appx: NMF_params}.

In our implementation, we used the \texttt{scikit-learn} module of Python \citep{scikit-learn}. The NMF algorithms are described in the \texttt{sklearn.decomposition.NMF} class, and the parameters are shown in appendix \ref{appx: NMF_params}. The procedure applied is illustrated in figure \ref{method}. First of all, we begin by replacing all non-numbers and negative values\footnote{Images we used contain some non-numbers (i.e., NaNs) because we masked out locations in the position-position-velocity cube of transitions where emission is not expected as described in \citet{Harada2024}. Negative values come from absorption lines that are present around GMC 5 for limited transition.} with 0 in the moment 0 images. Subsequently, we normalize the minimum value to be 0 and the maximum value to be 1, and make the two-dimensional data (512 pixels $\times$ 640 pixels) from a given transition one-dimensional ($512\times 640=327,680$ pixels). Then, we integrate all the emission line data into a matrix ($327,680 \mathrm{\>rows} \times 150 \mathrm{\>columns}$) by applying the same procedure to the 150 images. The data matrix \textsf{E} has entries $e_{ln}$ where $l$ is an index of a row with a range of $1 \leq l \leq L$ and $n$ is an index of a column with a range of $1 \leq n \leq N$. Here, $L\equiv l_\mathrm{max}=327,680$ and $N \equiv n_\mathrm{max}=150$. Applying NMF (non-negative matrix factorization) to this matrix, we decompose it into two matrices: the base matrix \textsf{B} with entries $b_{lm}$ with the range of $1 \leq m \leq M$ where $M\equiv m_\mathrm{max}=5$ and the coefficient matrix \textsf{C} with entries $c_{mn}$. Each column of the base matrix is then converted into a two-dimensional map, resulting in five maps defined as non-negative factor (NF) components (NF1 -- NF5). The order has no significance although we indexed NFs by number from NF1 to NF5. We identified trends in these maps and extracted emission lines similar to these NF components. The detailed structure of the NMF algorithm is shown in appendix 1.

% 5 is the number of components of NMF which we arbitrarily can choose it ranges from 0 to 150. 
In NMF, we can arbitrarily choose the number of components $M$ up to $\min(L,N)$ when the input matrix has the dimension of $L \times N$. Among these numbers, it is advisable to choose a proper number where we can interpret the data more effectively. If the number of components is too small, the error between $\bm{\textsf{E}}$ and $\bm{\textsf{BC}}$ increases. Conversely, the more components you have, the smaller the error becomes. However, physically meaningless components will be created, noisy structures start to appear, or some components become too localized (see the case of $M=8$ in appendix 3). \citet{Harada2024} discussed 5 components of PCA maintaining $\geq 96\> \%$ of the variance within the data. Following this, we decided to use 5 components in this study. This choice of the number of components is also justified by the reconstruction error between $\bm{\textsf{E}}$ and $\bm{\textsf{BC}}$ shown in appendix 2, where we see a steep decline of the specific error from $m \rightarrow m+1$ when $ m < 4 $.

\section{Results}\label{sec:results}
\subsection{Outline of each component}\label{subsec: outline}

\begin{figure*}[ht]
    \begin{center}
        \includegraphics[keepaspectratio, width=\linewidth]{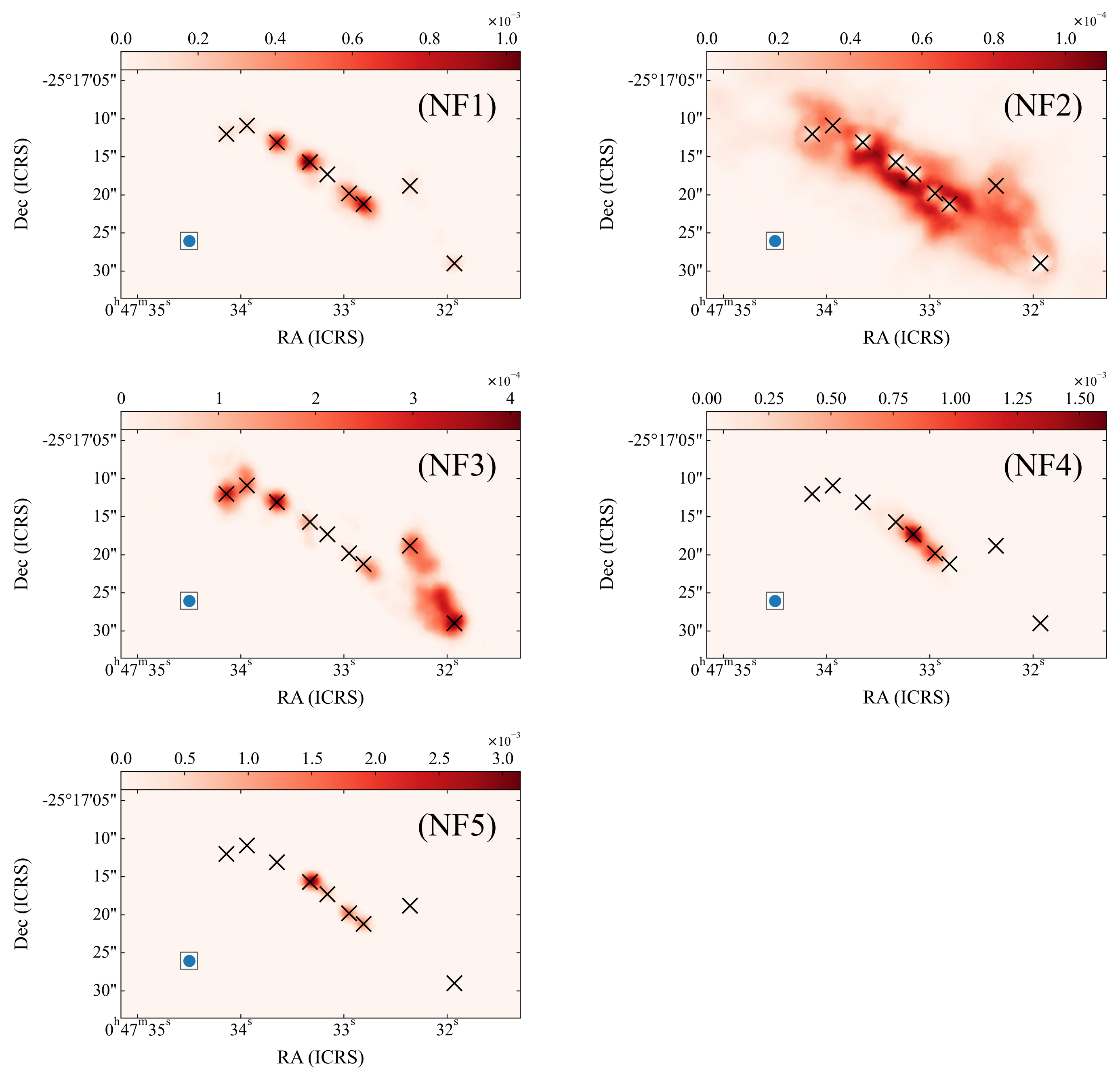}
    \end{center}
\caption{Maps of base matrices $\bm{\textsf{B}}$ (i.e., vectors $b_{l1}$, $b_{l2}$, $b_{l3}$, $b_{l4}$, and $b_{l5}$ transformed back to two-dimensional maps). Cross marks represent the positions of GMCs (table \ref{table:GMCs}). Circles on the lower left show the size of the beam which is equivalent to $1.6^{\mathrm{"}} = 27 \mathrm{\, pc}$.}
\label{fig:NF_map}
\end{figure*}

\begin{figure*}[ht]
    \begin{center}
        \includegraphics[keepaspectratio, width=\linewidth]{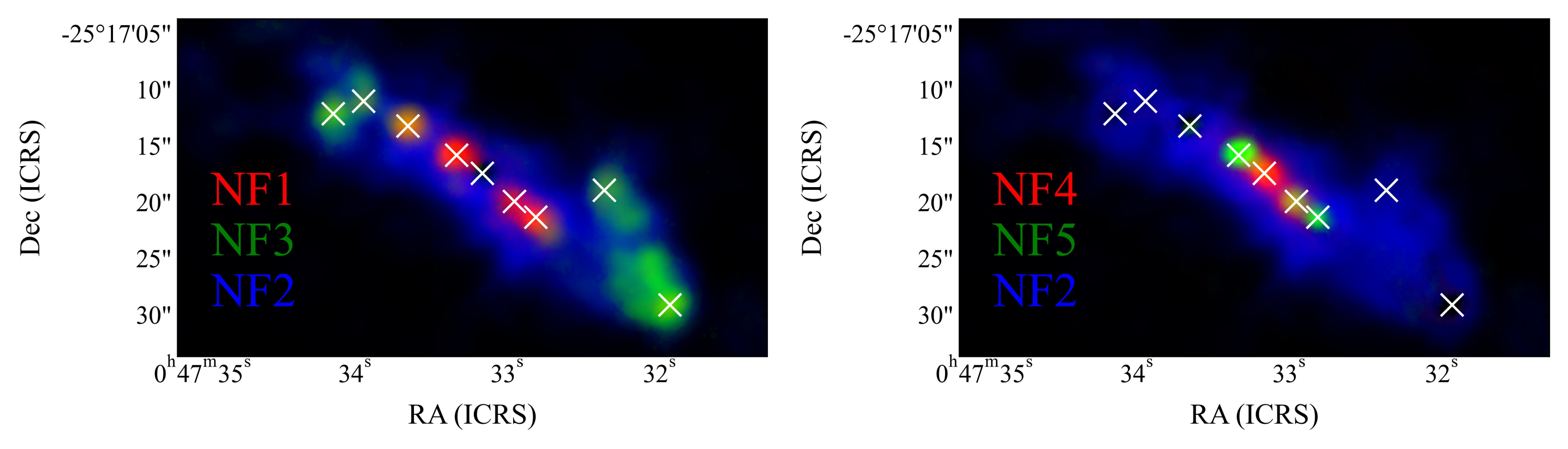}
    \end{center}
\caption{(left) The maps of NF1(red), NF3(green), NF2(blue) are shown as an RGB figure. (right) Same as the left panel, but with NF4(red), NF5(green), NF2(blue). We applied the method of \citet{Lupton2004}.}
\label{fig:NF_RGB}
\end{figure*}

\begin{figure*}[ht]
    \begin{center}
        \includegraphics[keepaspectratio, width=\linewidth]{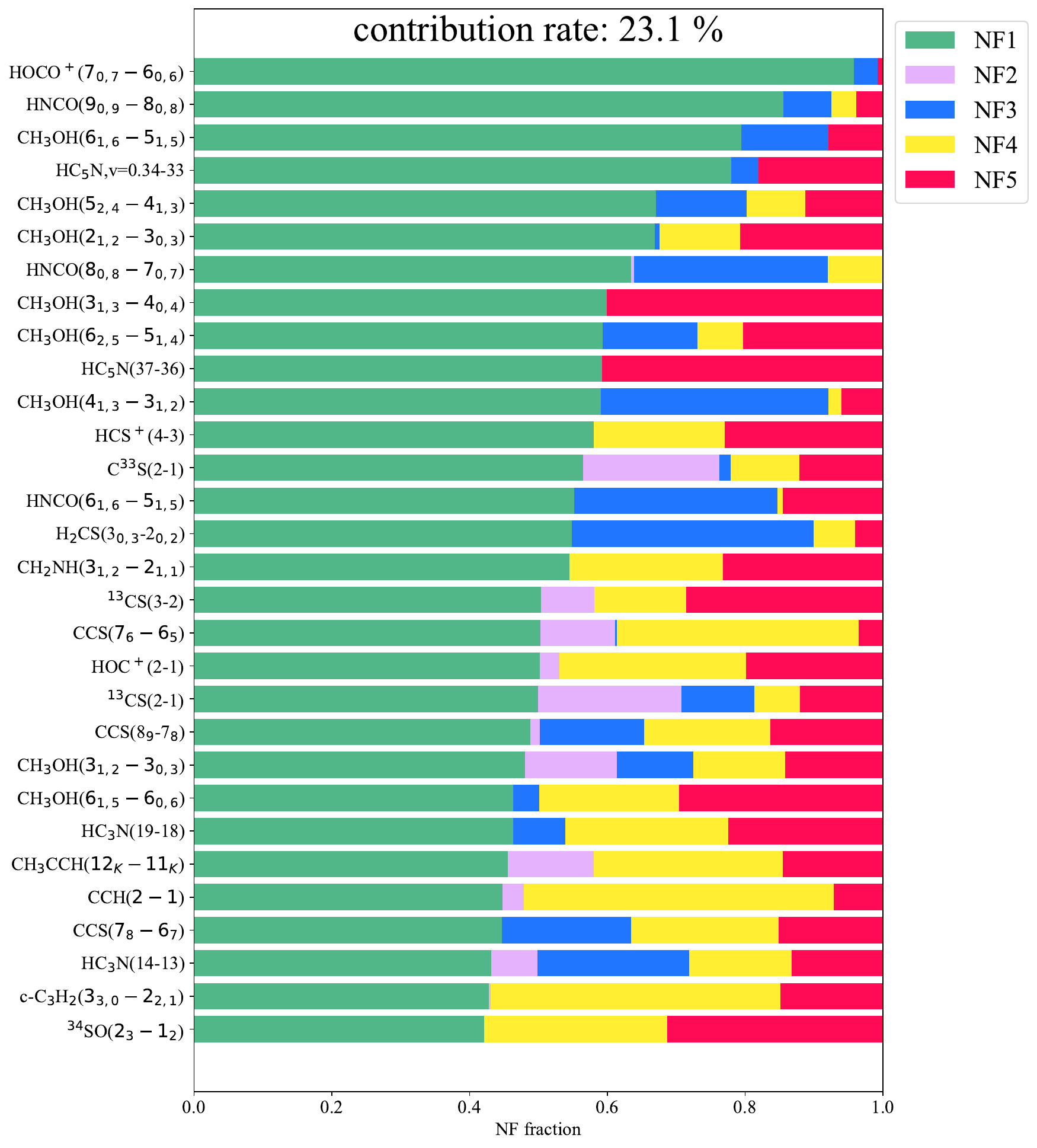}
    \end{center}
\caption{NF components for 30 selected transitions (see Section \ref{subsec: outline}), sorted according to the NF1 fractions relative to the sum of the NF1 to NF5 contributions.}
\label{fig:NF1}
\end{figure*}

\begin{figure*}[ht]
    \begin{center}
        \includegraphics[keepaspectratio, width=\linewidth]{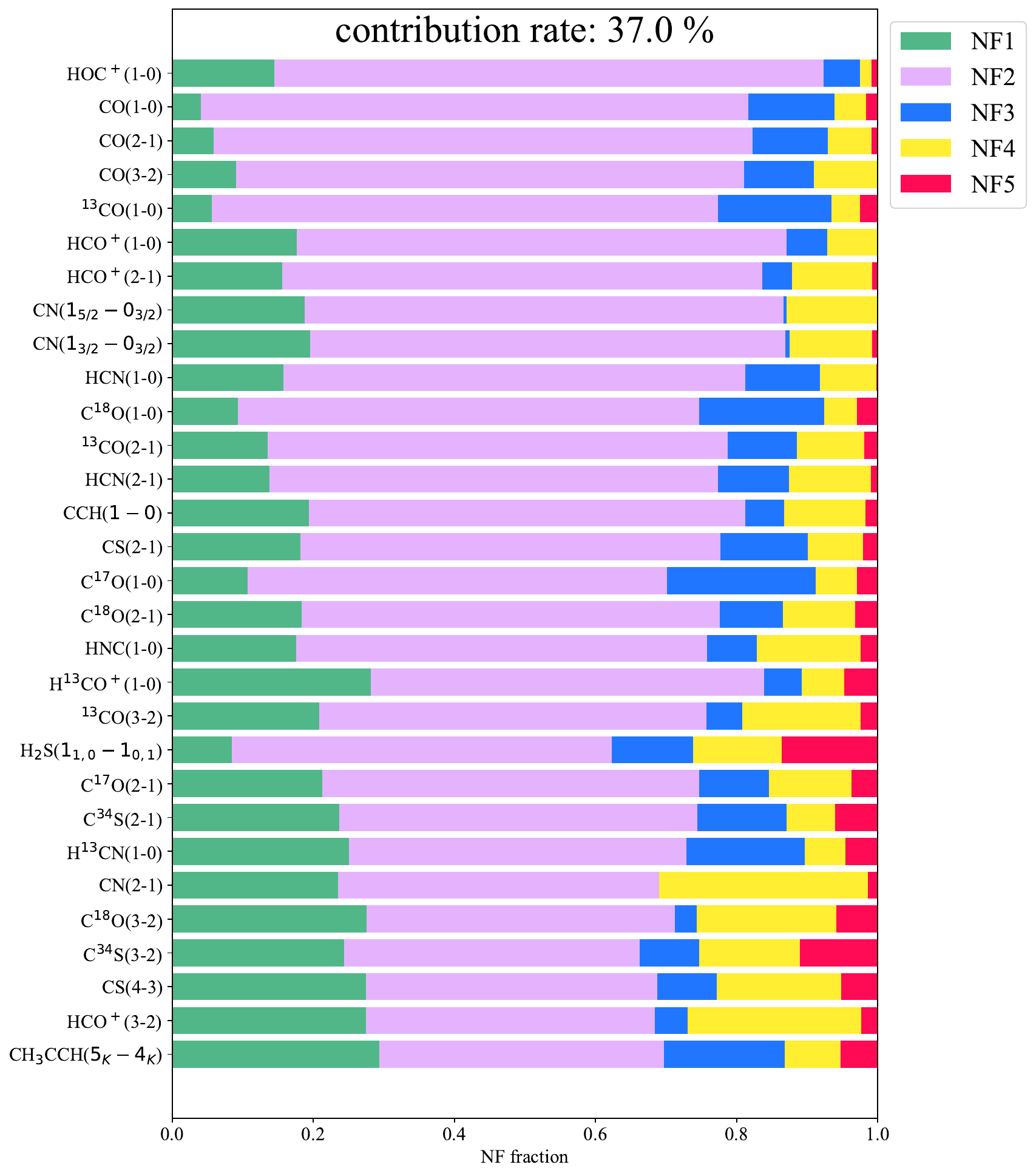}
    \end{center}
\caption{Same as figure \ref{fig:NF1}, but sorted by the NF2 fractions.}
\label{fig:NF2}
\end{figure*}

\begin{figure*}[ht]
    \begin{center}
        \includegraphics[keepaspectratio, width=\linewidth]{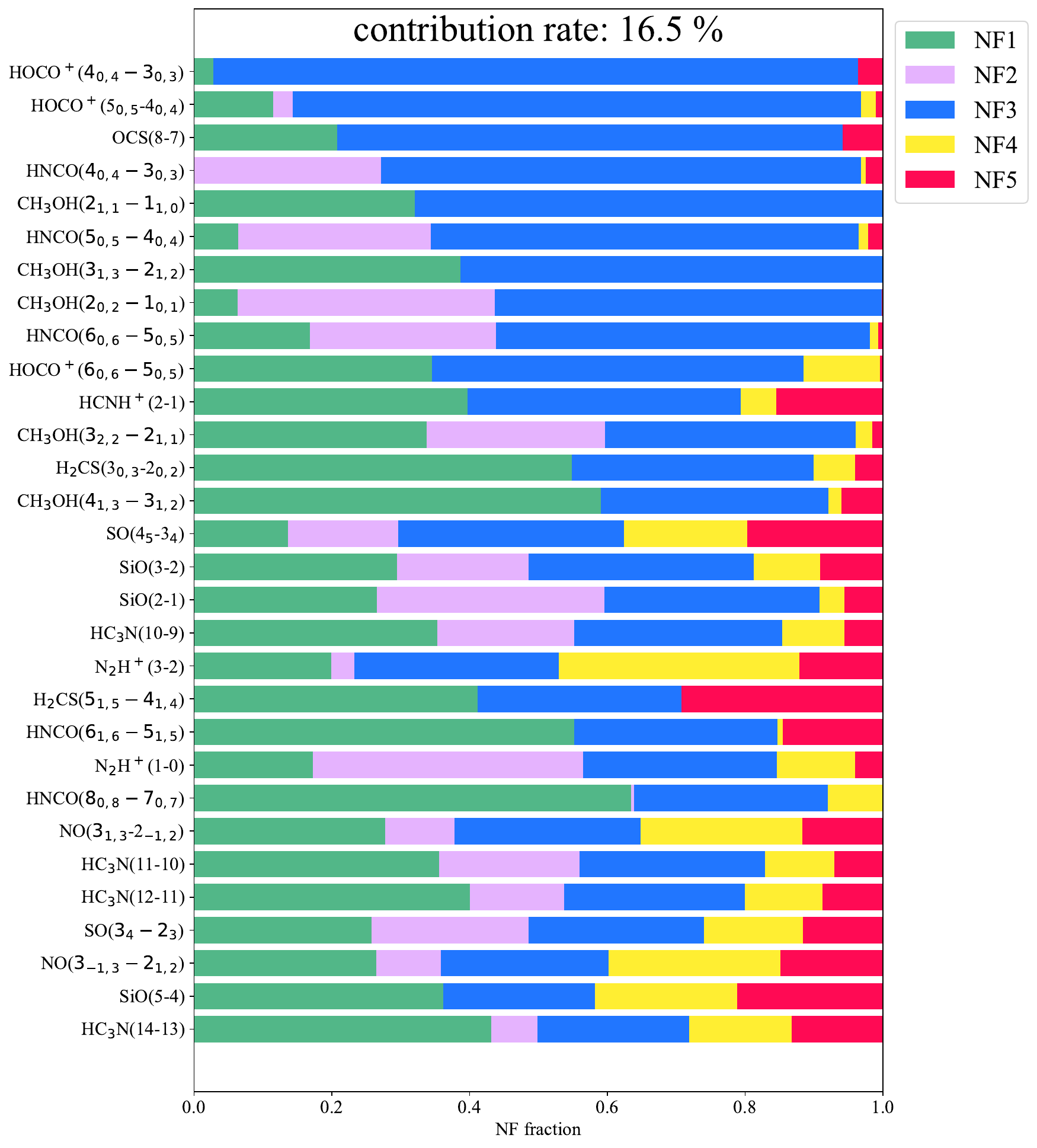}
    \end{center}
\caption{Same as figure \ref{fig:NF1}, but sorted by the NF3 fractions.}
\label{fig:NF3}
\end{figure*}

\begin{figure*}[ht]
    \begin{center}
        \includegraphics[keepaspectratio, width=\linewidth]{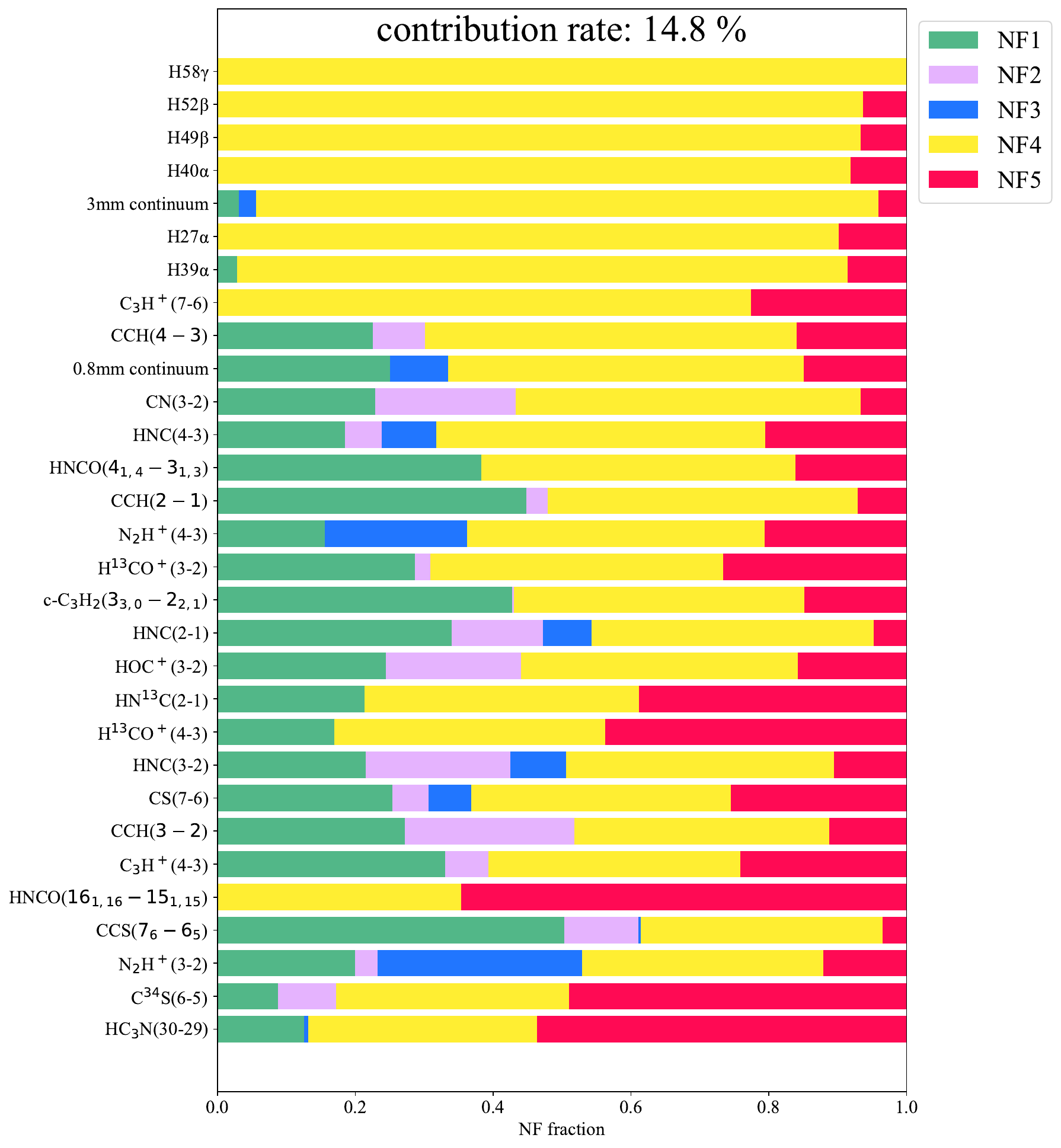}
    \end{center}
\caption{Same as figure \ref{fig:NF1}, but sorted by the NF4 fractions.}
\label{fig:NF4}
\end{figure*}

\begin{figure*}[ht]
    \begin{center}
        \includegraphics[keepaspectratio, width=\linewidth]{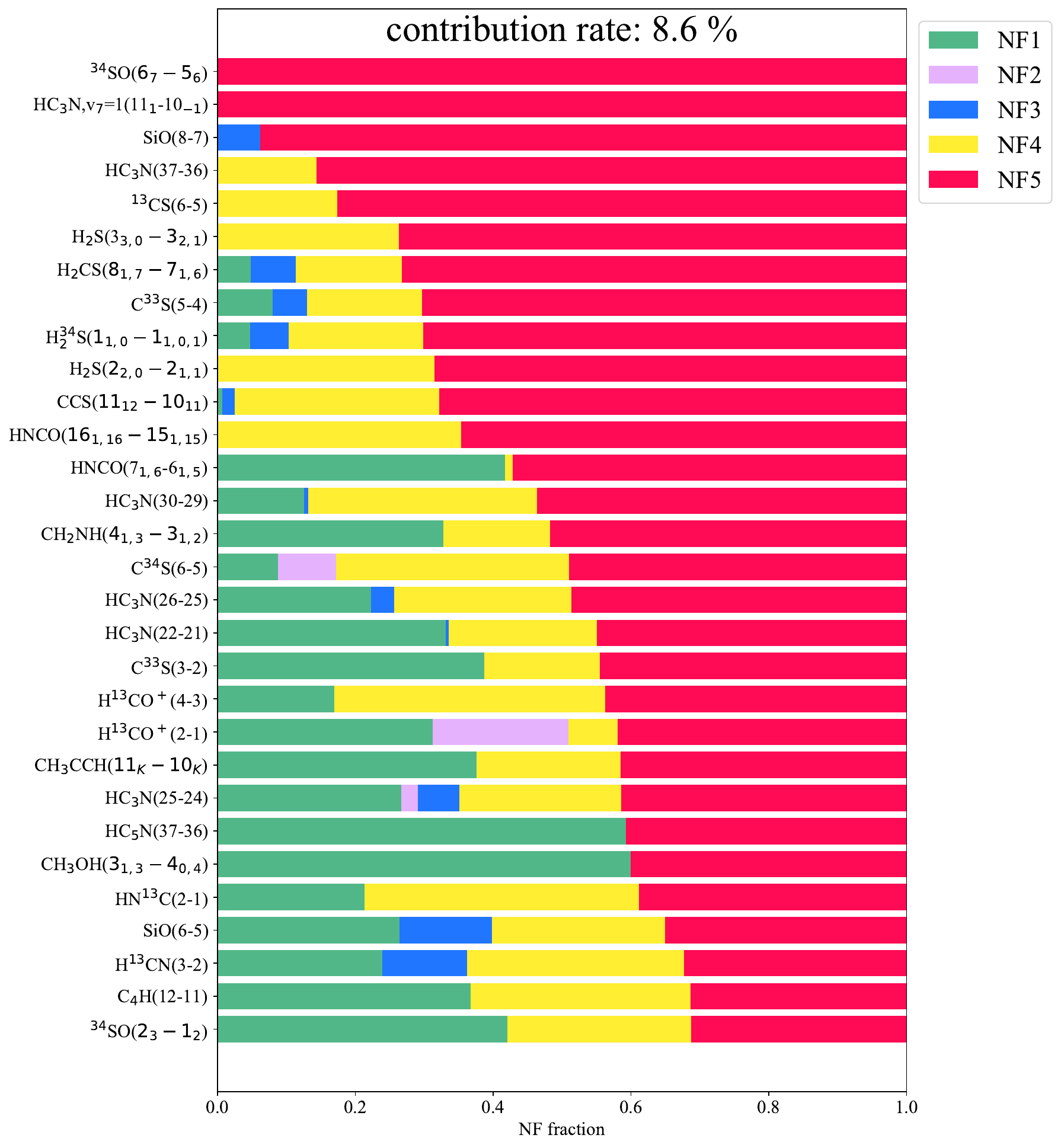}
    \end{center}
\caption{Same as figure \ref{fig:NF1}, but sorted by the NF5 fractions.}
\label{fig:NF5}
\end{figure*}

Five dimensionally reduced maps are shown in figure \ref{fig:NF_map}. Note that the order of each NF does not have any meaning. This is unlike PCA, where components are numbered in the descending order of variance explained.
%These maps are two-dimensional for each column of matrix $\bm{\textsf{B}}$ (base matrix) (figure \ref{method}).
These maps are constructed from each column of the base matrix $\bm{\textsf{B}}$ (i.e., one dimensional 327,680 pixels) back into the two-dimensional image of 512\>pixels\>$\times$\>640\>pixels (see figure \ref{method}). Cross marks in the figures explain GMC's positions in the center of NGC~253 (table \ref{table:GMCs}). Figure \ref{fig:NF_RGB} is the red-green-blue (RGB) composite image of NF maps. It contains the same information as figure \ref{fig:NF_map}, but we show this figure to highlight the contrast between different components. Here, we show NF1(red), NF3(green), and NF2(blue) in the left panel, NF4(red), NF5(green), and NF2(blue) in the right panel.

From the coefficient matrix $\bm{\textsf{C}}$, we calculate a quantity called NF fraction (defined below) shown in figures \ref{fig:NF1} -- \ref{fig:NF5}. %These figures represent NF components of each emission line in matrix $\bm{\textsf{C}}$ after applying NMF. 
Each NF fraction corresponds to {\it a degree of similarity} to each NF map. Here we define the NF$x$ fraction of a transition $n$ to be \[c_{xn}/\sum^{5}_{m=1}c_{mn}\] 
and the contribution rate of NF$x$ to be \[ \sum_{i=1}^{N} c_{xi} / \sum_{i=1}^{M} \sum_{j=1}^{N} c_{ij}. \] 
Because we scaled the base matrix so that the sum of each column is 1, the contribution rates are the direct measures of how much flux in input images each NF contributes to the total flux of input images (see appendix 1).

We show transitions with the top 30 NF fractions for each $m$ in these figures, and describe the contribution rate at the top of each figure.

\begin{itemize}
    \item NF1\\
    The base matrix map (figure \ref{fig:NF_map} top left) shows that the values in GMCs 3, 4, 6, and 7 are high. Small values exist in GMC 5 and outside GMCs. Characteristic transition lines of NF1 (i.e., transitions that have high NF1 fraction) are HOCO$^+$(7$_{0,7}-6_{0,6}$), HNCO($9_{0,9}-8_{0,8}$), or CH$_3$OH ($6_{1,6}-5_{1,5}$) (figure \ref{fig:NF1}). The above transitions have excitation temperatures of ~30-60 K, which are neither low nor high.
    %These species are known as weak shock tracers.
    \item NF2\\
    This NF component has some structures extending out from the galactic plane (figure \ref{fig:NF_map} top right). The positions of some of the GMCs appear as cavities in the NF2 map. The maximum pixel value of NF2 is the lowest in all NF maps. Carbon monoxide, its isotopologues (e.g., $^{13}$CO), and HCO$^+$ have high NF2 fraction. This extended structure may result from outflows driven by intense star formation in NGC\,253 as reported by \citet{Krieger2019}. The ionized gas is blowing out from the central part of NGC\,253 with a cone of opening angle 60$^\circ$, which is generally observed with X-rays (e.g., \cite{Weaver2002}). The molecular gas which is likely the flow entrained by the ionized outflow exists in the wall of the outflow cone \citep{Walter2017}. In the ALCHEMI data, the molecular outflow may be seen in the region coming out of GMCs 2, 3, 4, and 7.
    \item NF3\\
    There are peaks at the outskirts of the CMZ (GMCs 1, 2, 7, 8, and 9), and there are lower but some contributions closer to the kinematic center at GMCs 3 and 6. Characteristic transition lines are low energy transitions of CH$_3$OH, HNCO, HOCO$^+$, and OCS. Although SiO is also known to be a shock tracer, the highest NF3 fraction of SiO (strong shock tracer) overall NF3 fraction is 0.33. This fraction is relatively low compared to 0.61 -- 0.94 for the low-excitation transitions of other shock tracers. Among shock tracers, the mechanism of the SiO enhancement is different from that of other species. A significant increase of SiO is likely caused by sputtering due to fast shocks ($>20\,$km s$^{-1}$) of Si from dust grains \citep{Schilke1997}. Meanwhile, the enhancement of other species is due to the sputtering of ice, which requires slower shocks. The difference in the NF3 fractions may result from the strength of shocks.
    \item NF4\\
    The NF4 component is mostly concentrated in compact regions around GMCs 4 and 5. We find that GMC 5 has a higher value and larger structure than that of GMC 4. The transition lines with the high NF4 fraction are the radio recombination lines (RRLs; H39$\alpha$, H$\beta$, and H$\gamma$), the UV tracers C$_3$H$^+$ and CCH.
    \item NF5\\
    The NF5 component has compact structures in GMC 3, 4, and 6. Among them, GMC 6 has the highest value, while GMC 3 and 4 are comparable. The excited lines with high NF5 fractions are the vibrationally excited lines (HC$_3$N,v$_7$=1(11$_1$-10$_{-1}$)) or high-energy level of transition, such as SO, SiO, HC$_3$N, etc.
\end{itemize}

\subsection{Energy dependence of the NF components}\label{subsec:Energy dependence of NF component}
\begin{figure*}[ht]
    \begin{center}
        \includegraphics[keepaspectratio, scale=1.4]{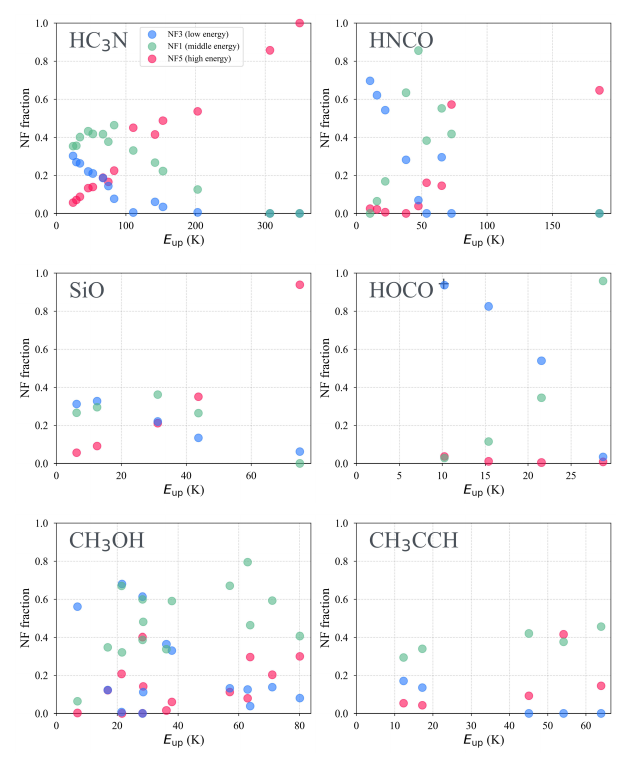}
    \end{center}
\caption{NF1 (blue, low energy), NF3 (green, middle energy), and NF5 (red, high energy) of four molecular species are shown. These six reference molecular species are HC$_3$N (top left), HNCO (top right), SiO (middle left), HOCO$^+$ (middle right), CH$_3$OH (bottom left), and CH$_3$CCH (bottom right). The horizontal axis represents the upper-state energy ($E_\mathrm{up}$) and the vertical axis represents the NF fraction.}
\label{fig:NF1map}
\end{figure*}

We analyze how NF components depend on the excitation of molecular transitions, using upper-state energy ($E_\mathrm{up}$) as an indicator. The NF1, NF3, and NF5 fractions of HC$_3$N, HNCO, SiO, HOCO$^+$, CH$_3$OH and CH$_3$CCH are shown in figure \ref{fig:NF1map}. 
The low-energy transition lines of these molecular species except for CH$_3$OH and CH$_3$CCH have the high NF3 fractions, while the high-energy ones have the high NF5 fractions. The transitions with $E_\mathrm{up}$ in between have the high NF3 fractions. As the energy level increases, the NF1 and NF3 fractions decrease while the NF5 fractions increase.
%Highly excited transitions of molecular species having high NF1 component is occupied by the NF5 value.
Similar trends can be seen in almost every molecular species with a wide $E_\mathrm{up}$ range. On the other hand, a few molecular species such as CH$_3$CCH and CH$_3$OH have different trends. As for CH$_3$CCH($11_K-10_K$)\footnote{As described in \citet{Harada2024}, all the $K$-ladder transitions are blended, and we treat them as one transition. Judging from the spectral shape, the $K=0$ transition gives the highest contribution to the intensity.}, the second highest temperature transition line of CH$_3$CCH, has an NF5 fraction of about 0.4, which is higher than other transition lines. When it comes to methanol (CH$_3$OH), these transition lines do not show the clear trend described above.
%When it comes to methanol, high-excitation transitions have the NF5 fractions of around 0.6 (CH$_3$OH($3_{1,3}-4_{0,4}$)), 0.4 (CH$_3$OH($2_{1,2}-3_{0,3}$)), or 0.3 (CH$_3$OH($3_{1,2}-3_{0,3}$)) and (CH$_3$OH($1_{1,0}-1_{0,1}$)) despite the low excitation transition line.
This lack of trend among the excitation temperatures and NF fractions of methanol may be attributed to non-thermal excitation such as masers described by \citet{Humire2022}.

\section{Discussion}\label{sec:discussion}
\subsection{Interpretation of each NF component}\label{subsec:Interpretation of each NF component}
From NF1 to NF5, each NF map shows a distinct structure. We were able to identify each NF component involved in star formation. Furthermore, emission lines that describe the structure involved in star formation are defined as tracers.
\begin{itemize}
    \item NF1 \par
    We propose that the NF1 structure represents the distribution of middle-excited shock tracers because NF1 contains a high ratio of HOCO$^+$(7$_{0,7}-6_{0,6}$), HNCO($9_{0,9}-8_{0,8}$), and CH$_3$OH($6_{1,6}-5_{1,5}$). As seen in Section \ref{subsec:Energy dependence of NF component}, this trend is best seen in combination with NFs 3 and 5.
    \item NF2 \par
    We argue that NF2 represents the low-density gas spreading throughout the galaxy. The NF2 map shows a porous structure that avoids locations of GMCs, which likely have higher densities. It also contains streamers extending vertically from the galactic plane, including outflows detected by \citet{Bolatto2013}. \citet{Tanaka2024} also found that some vertically extended streamers have high temperatures, suggesting that these regions have been affected by energetic events. It is consistent with the idea that these are outflows. If NF2 traces diffuse gas, it is quite understandable that transitions of $^{12}$CO have the high NF2 fraction because of the relatively low critical densities and high intensities. However, HOC$^+$(1--0) has the highest NF2 fraction. We attribute this trend to the ample presence of strong ionizing sources such as UV photons or cosmic rays in NGC\,253 \citep{Harada2021}.
    \item NF3 \par
    NF3 likely traces the shock distribution. The region where NF3 is prominent (GMC 1, 2, 7, 8, 9) is 150\,pc -- 350\,pc from the galactic center (the TH2 region, \cite{Turner1985}; near GMC 5). This distribution is similar to the position of Class I masers, which are usually associated with shocks \citep{Ellingsen2017,Chen2018,Gorski2019,Humire2022}. Additionally, \citet{Huang2023} determined the location of shocks in NGC\,253 with SiO and HNCO as shock tracers. They reported results that are also consistent with this distribution. We also consider the fact that familiar shock tracers such as HNCO \citep{Meier2005}, CH$_3$OH, HOCO$^+$ \citep{Harada2022} and OCS have high NF3 rates as evidence that NF3 traces shocks in the galaxy. However, the rate of SiO, which is a well-known strong shock tracer is relatively low compared to other shock tracers. This result may indicate that the shock caused a large amount of sublimation of ice on dust. Although a small amount of SiO or elemental Si may be present in ice, a significant increase of SiO occurs when silicon is sputtered out from the dust core. 
    \item NF4 \par
    The distribution of the NF4 map resembles that of the SFR traced by RRLs and the 3\,mm continuum. \citet{Bendo2013} identified the location of starburst regions based on 99.02\,GHz free-free emission and H40$\alpha$ emission in the central region of NGC\,253. Our distribution of NF4 is almost identical to that found by this study. The fact that the RRLs and the 3\,mm continuum contain an NF4 fraction of $\geq 0.89$, and the 0.8\,mm continuum contains about 0.5, is also consistent with this possibility.
    \item NF5 \par
    The NF5 structure is thought to represent young star-forming regions. As previously stated, GMC 6 has the highest values. \citet{Rico2020} focused on 14 super star clusters (SSCs) in the central part of NGC\,253 in the $0\farcs3$ (=5\,pc) resolution data and determined the age of the SSCs from the luminosity of the SSCs and the protostars. The location of the youngest SSCs and the NF5 map are generally consistent. The emission lines with high percentages of NF5 have relatively high upper-state energy ($E_\mathrm{up}$), such as vibrationally-excited lines and highly excited lines. It is thought that infrared radiation emitted from protostars heats the molecular gas and excites many species to comparatively high energy levels.
\end{itemize}

It is important to note that there are some caveats in using NMF. For example, \citet{Mijolla2024} found that NMF can group multiple components of similar properties even if these components have different properties. 

\subsection{Comparison with PCA}
We discuss the advantages and disadvantages of NMF when compared with PCA \citep{Harada2024}. Note that \citet{Harada2024} used hexagonal pixels that are larger than the ones of the original images, while we show images using the pixels of original images as we did for NMF. Results using two different pixel binning methods are essentially the same.  First, we show a brief overview of PCA (see \cite{jolliffe2002} and others for details).

\begin{figure*}[ht]
    \begin{center}
    \includegraphics[keepaspectratio, width=\linewidth]{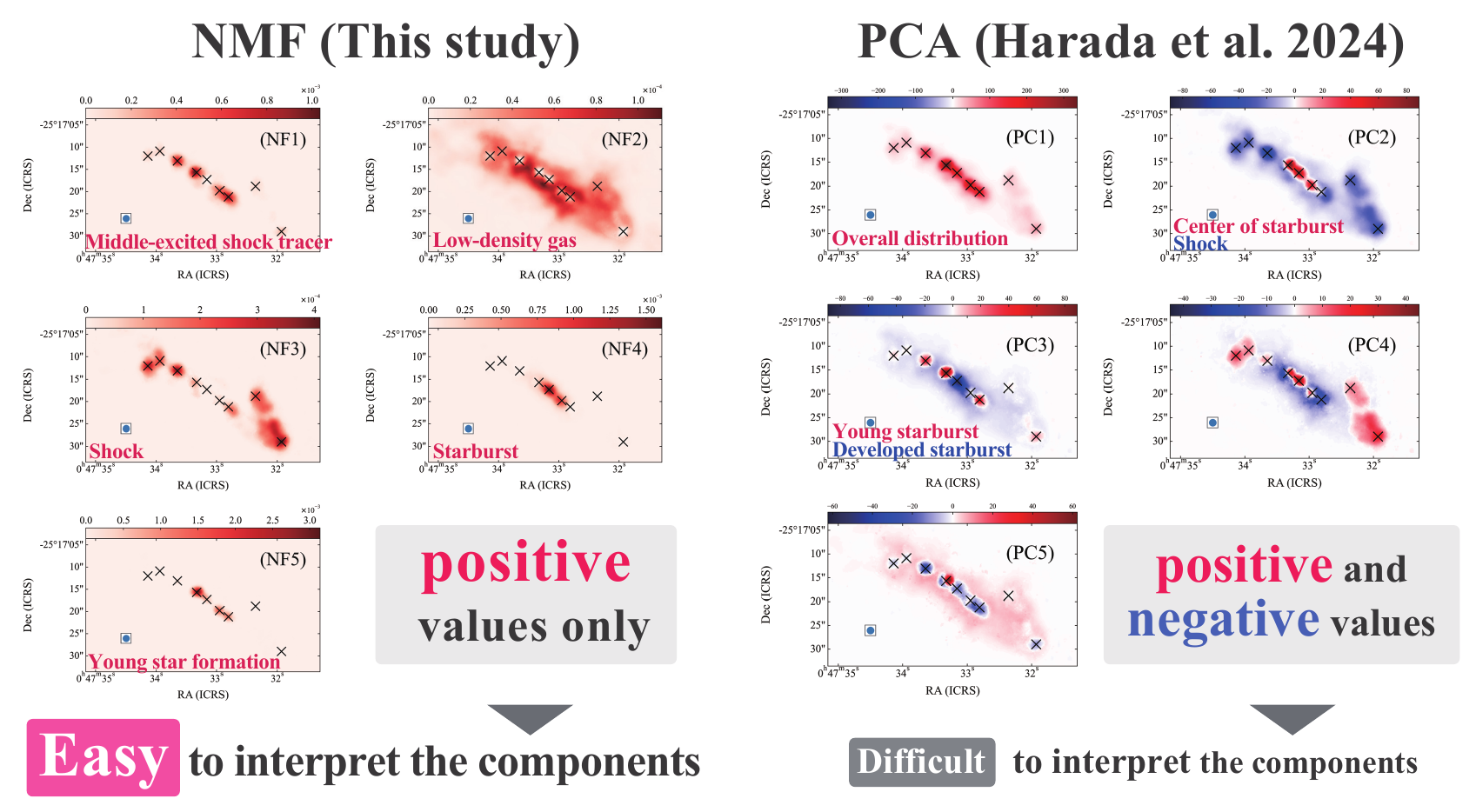}
    \end{center}
\caption{The results of NMF (this study) and PCA \citep{Harada2024}. In the PCA component maps, the red area represents positive and the blue area represents negative values. Physical interpretations are provided with red (positive) and blue (negative) texts on each map.}
\label{fig:PCA_comparison}
\end{figure*}

In PCA, it is common to first standardize measurements within each variable so that they have a mean of zero and a standard deviation of one. In our case, variables are different transitions and measurements are intensities of each transition from all the pixels. In the data space with transition intensities as axes, data points are the standardized intensities from all the pixels. PCA sets up a new orthogonal coordinate system in this data space. The direction of the largest variance becomes the first principal component (PC), PC1. The second component (PC2) is chosen in the direction of the second largest variance that is orthogonal to PC1. Similarly, a subsequent component lies in the direction orthogonal to prior PCs with the next highest variance. If data are somewhat correlated, one can explain most of the variance in the data with a far fewer number of principal components than a number of transitions. This means the dimensionality is reduced. Maps shown in figure \ref{fig:PCA_comparison} (right) are obtained by projecting data points onto the PC axes, i.e., maps of PC scores.

One of the main differences between PCA and NMF is that PCs are derived on the order of variances in the data while there is no preferred order in NFs. As discussed later, the percentage of variance explained by PC1 is very large compared with other PCs (79.4$\>\%$ for PC1, 9.0$\>\%$ for PC2, 5.0$\>\%$, $\cdots$). However, the difference in contributions among NFs is smaller than that of PC (contribution rate of 23.1$\>\%$ for NF1, 37.0$\>\%$ for NF2, 16.5$\>\%$ for NF3, 14.8$\>\%$ for NF4, and 8.6$\>\%$ for NF5). Further, the contribution rate tends to be high for an NF with an extended structure because input images are normalized to have maximum values of 1, and images with extended emission have more flux. Considering the larger contribution of extended emission, the difference in importance among NFs is small. Another difference is that PCs are orthogonal to each other, while NMF does not require orthogonality between the components. These differences in the dimensionality reduction method lead to the differences in the components extracted.

\citet{Harada2024} analyzed exactly the same NGC\,253 transition lines and continuum data using PCA. In figure \ref{fig:PCA_comparison}, we compare the maps obtained from PCA and NMF. We find that there are both similarities and differences between the results obtained from these methods. \citet{Harada2024} concluded that PC1 represents the entire molecular gas distribution. It explains much of the ALCHEMI data because the percentage of variance explained is 79.4$\>\%$. PC1 is the prominent component of PCA, but NMF could not extract the component corresponding to PC1. This difference originates from the fact that PCA extracts a component that explains the most variance in data as PC1 while the degree to which each component explains the data is similar to each other among NF components.

PC2 (with a contribution of 9.0$\>\%$) shows different characteristics depending on the positive and negative values, with the positive ones possibly hinting at the central starburst region and the negative may represent the distribution of shocked and low-excitation regions. PC2 has the trend related to the $E_\mathrm{up}$. Positive PC2 is similar to NF4 and NF5 with large values of PC scores in GMC 5 and 6, although none of the PC2 positives are fully consistent with any of the NMF components. Indeed, the transition lines characteristic of PC2 are RRLs, H$_2$S, and high-energy levels of HC$_3$N, which have a high ratio of NF4 or NF5. Thus, the positive PC2 has the morphology of a combination of NF4 and NF5. On the other hand, the negative PC2 was considered consistent with NF3, which has a lower upper-state energy and is considered to represent a shock region (see section \ref{subsec:Interpretation of each NF component}). In terms of shocked regions, similar components could be extracted in PCA and NMF, although NMF derives components more individually. Low-excited transitions of HNCO, HOCO$^+$, and CH$_3$OH are shock tracers in PCA as well. 

The difference in the evolutionary stages of the starburst appears in  PC3, with positive PC3 (with a contribution of 5.0$\>\%$) representing young star formation and negative PC3 representing developed star formation. In the components of NMF, NF5 may represent young star formation (see section \ref{subsec:Interpretation of each NF component}), which is consistent with some of the positive PC3. The NF4 and the negative distribution of PC3 are very similar; molecular transition lines with a high positive fraction of PC3 are almost identical to those with a high fraction of NF5, such as SiO(8-7), vibrationally excited lines (HC$_3$N,v$_7$=1(11$_1$-10$_{-1}$)), and cyanopolyynes in their vibrational ground state (HC$_3$N, HC$_5$N). Molecular transition lines with a high negative fraction of PC3 are RRLs and 3\>mm continuum, which are also consistent with emission lines with a high NF4 fraction. We also note that NMF is better at tracing young starbursts. While the positive PC3 score is seen in GMC7, there is no active massive star formation in this GMC. NF5 more accurately traces young starbursts by having strong features only at GMCs 3, 4, and 6.

The most different result between PCA and NMF applied to the ALCHEMI data seems to be the diffuse gas extended through the galaxy, which is well extracted in NF2 using NMF. On the other hand, the PCA results are less pronounced, appearing only as the negative PC4 (1.8$\>\%$) and the positive PC5 (1.2$\>\%$). On the other hand, a component corresponding to NF1, which is thought to be a distribution of middle-excited shock tracers, is absent in PCA.

\begin{figure*}[ht]
    \begin{center}
    \includegraphics[keepaspectratio, width=\linewidth]{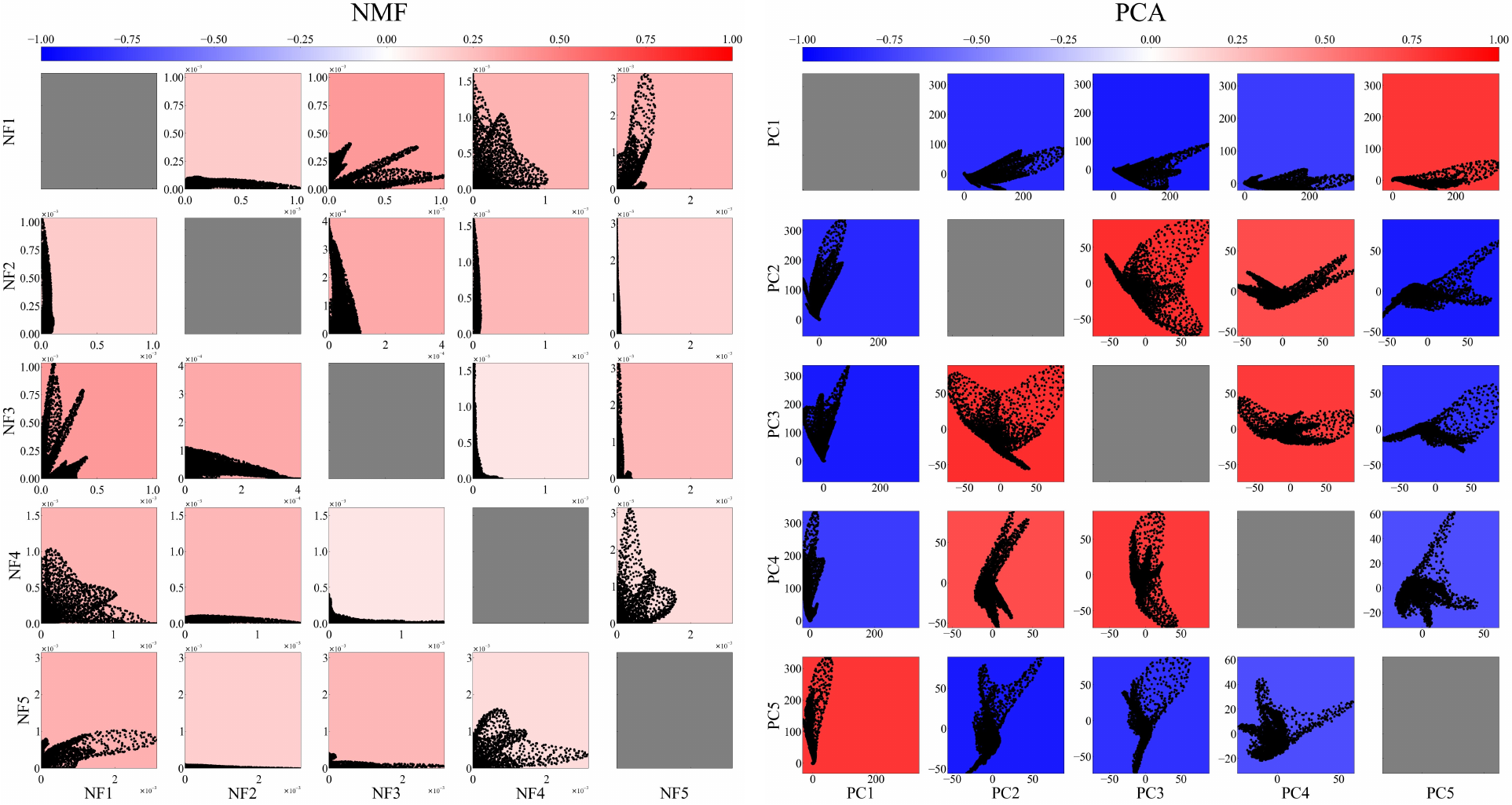}
    \end{center}
\caption{(Left) Pixel-by-pixel values in NF maps for NFs1--5 are plotted against each other. (Right) Same as the left panel, but PC scores for PCs 1--5. For both panels, the background color of each figure represents the value of Spearman’s rank correlation coefficients. The darker the color, the closer the absolute value is to 1.0 (i.e., the stronger the monotonic relation.)}
\label{fig:NFPC_relation}
\end{figure*}
Here we discuss the quantitative difference between NMF and PCA with Spearman’s rank correlation coefficients \citep{Spearman1904}. Figure \ref{fig:NFPC_relation} shows the pixel-to-pixel relation between the NF maps (left panel) and PC scores (right panel). The darkness of the background color in each figure shows the magnitude of Spearman’s rank correlation coefficients ($r_\mathrm{s}$). The Spearman's rank correlation coefficient is a useful indicator of how monotonically related the two variables are. We choose this coefficient over the Pearson correlation coefficient because Pearson correlation coefficients become zero for PCA because PC scores are orthogonal to each other, but that does not mean that PCs are not tracing the same physical component. Figure \ref{fig:NFPC_relation} shows that results of PCA have larger absolute values of $r_\mathrm{s}$ (with $|r_\mathrm{s}|\sim \,$1.0), indicating that there is a strong correlation between each PC. On the other hand, each NF component does not have a strong correlation with each other. This means that NMF is a more advantageous method than PCA for extracting uncorrelated and independent components. 

The above results suggest that the strength of PCA is in extracting the most significant factors that determine the data variation, in our case, the overall molecular distribution and excitation. On the other hand, NMF extracts highly independent components whether they are diffuse components as shown in NF2 or local structures shown in other components. This suggests that NMF is good at extracting different physical processes that are localized or diffuse, but not affected by the overall distribution of the molecule. This is an observational confirmation that NMF is effective in the interpretation of astronomical data, in addition to a theoretical work that previously supports this conclusion \citep{Mijolla2024}.

% \citet{Harada2024} analyze the same NGC\,253 data as in this study using principal component analysis (PCA). They used the same emission lines as we did for analysis. PCA is a well-known method of dimensionality reduction, similar to NMF. The components of this method differ in the degree to which they can explain the initial data. They show that PC1 (79.4$\>\%$ of all data explained) represents the distribution of overall molecular gas. PC2(9.0$\>\%$) shows different features according to positive and negative. They thought that  PC2 positive traces the central starburst region, contrary to this, PC2 negative traces shocks or cloud collisions. It was suggested that PC2 also has to do with upper-state energy ($E_\mathrm{up}$). Difference in evolutionary stages of starburst is determined by PC3. They concluded PC3 (5.0$\>\%$) positive is relative to young star formation, their negative is well-developed star formation. The most different feature between PCA and NMF was the diffuse gas spreading across the galaxy. It was efficiently detected using NMF (NF2). On the other hand, the result of PCA is not remarkable to the extent that it appears in PC4 (1.8$\>\%$) and PC5 (1.2$\>\%$). We consider that NMF may be good at extracting diluted components.

\subsection{Useful tracers to understand the physical state}
\begin{figure*}[ht]
    \begin{center}
    \includegraphics[keepaspectratio, width=\linewidth]{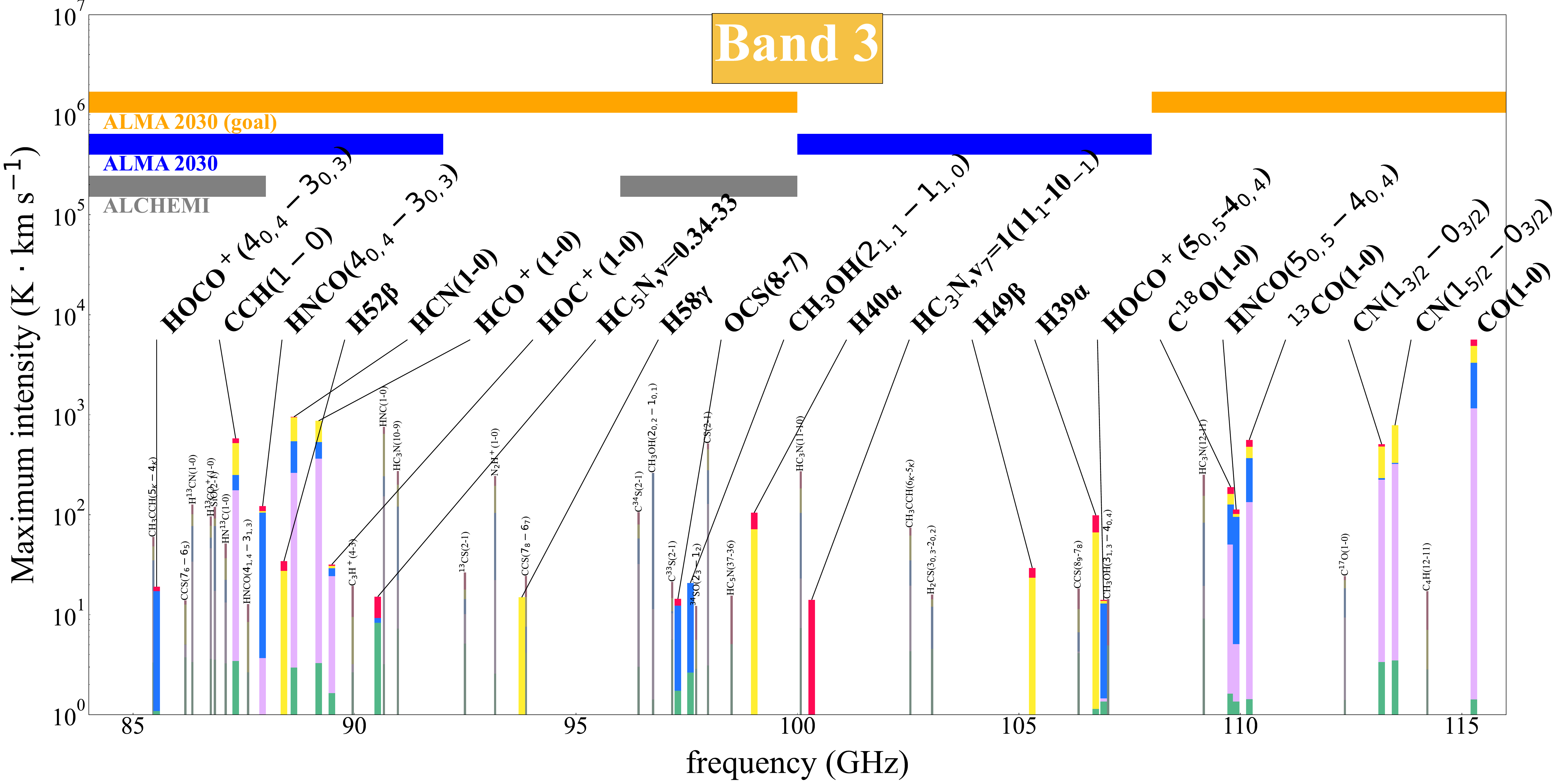}
    \end{center}
\caption{Transitions used in our study in Band 3 with NF fractions higher than 0.6. The color of each lines represents the same as figure \ref{fig:NF1} to \ref{fig:NF5}. For Band 3, the upgraded receiver will be Band 2+3, instead of Band 3 alone. Please note that we only show transitions used in our study and exclude transitions with some blending.}
\label{fig:Useful_tracers_band3}
\end{figure*}

\begin{figure*}[ht]
    \begin{center}
    \includegraphics[keepaspectratio, width=\linewidth]{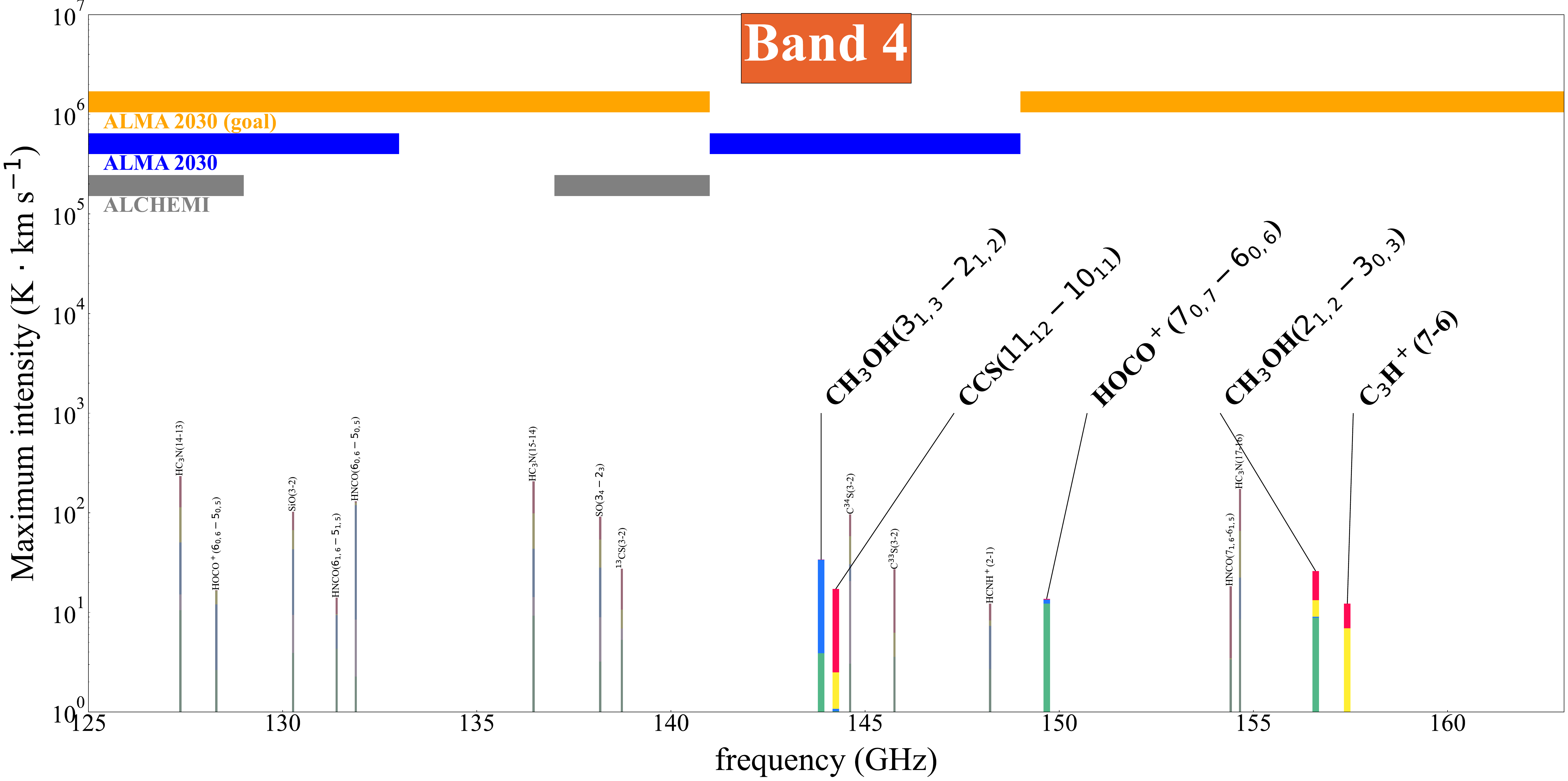}
    \end{center}
\caption{Same as figure \ref{fig:Useful_tracers_band3}, but for Band 4.}
\label{fig:Useful_tracers_band4}
\end{figure*}

\begin{figure*}[ht]
    \begin{center}
    \includegraphics[keepaspectratio, width=\linewidth]{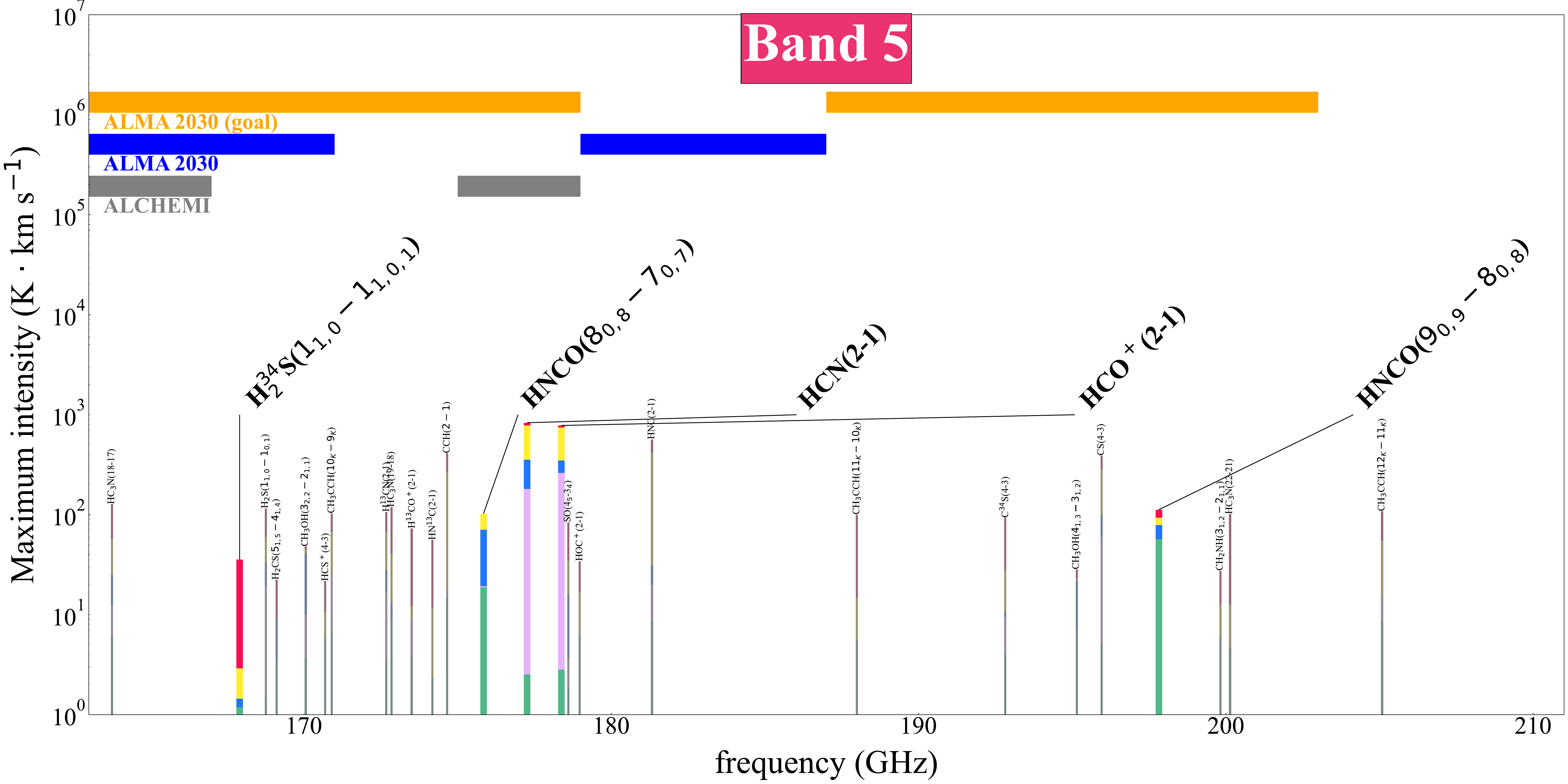}
    \end{center}
\caption{Same as figure \ref{fig:Useful_tracers_band3}, but for Band 5.}
\label{fig:Useful_tracers_band5}
\end{figure*}

\begin{figure*}[ht]
    \begin{center}
    \includegraphics[keepaspectratio, width=\linewidth]{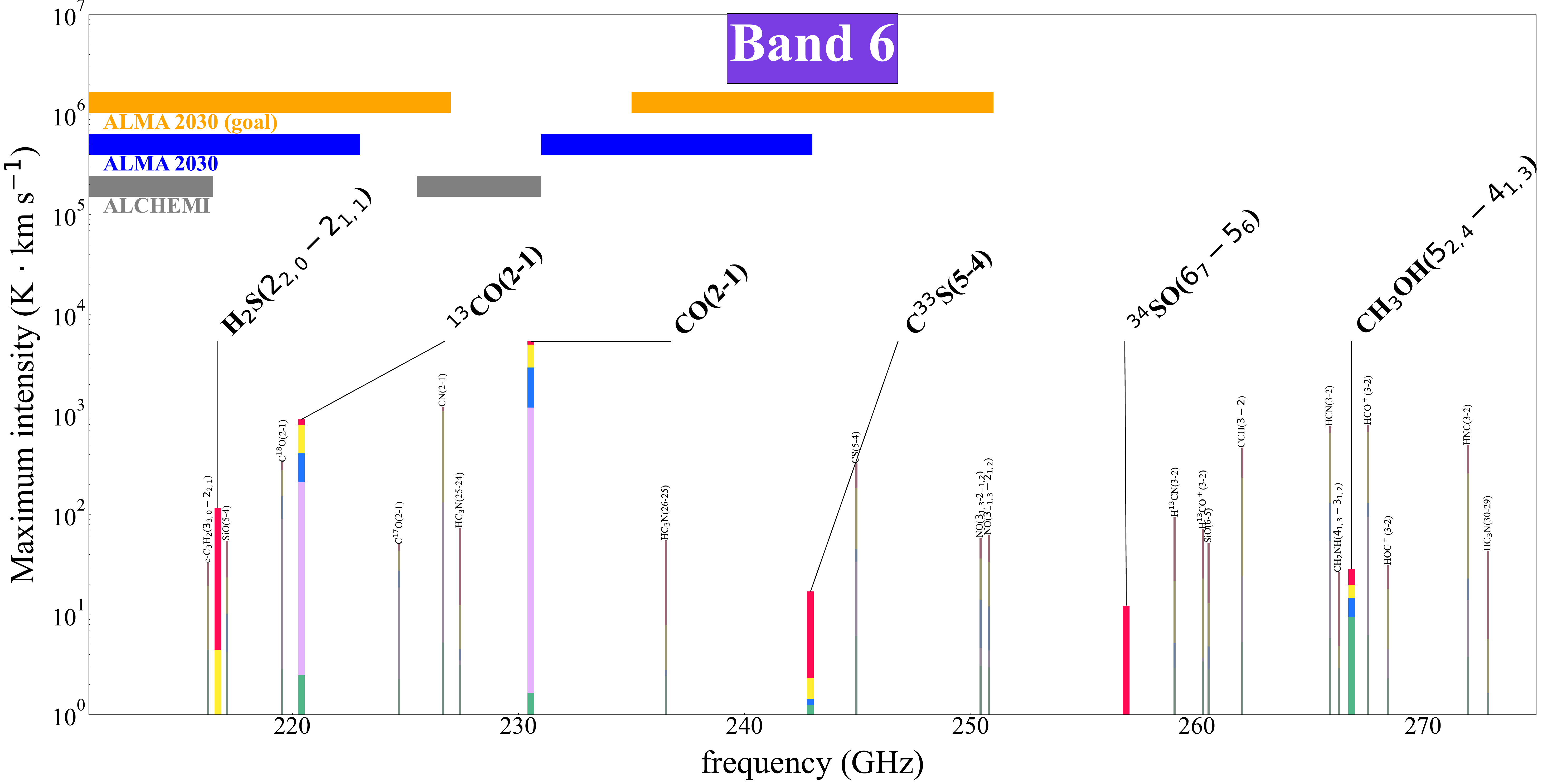}
    \end{center}
\caption{Same as figure \ref{fig:Useful_tracers_band3}, but for Band 6.}
\label{fig:Useful_tracers_band6}
\end{figure*}

\begin{figure*}[ht]
    \begin{center}
    \includegraphics[keepaspectratio, width=\linewidth]{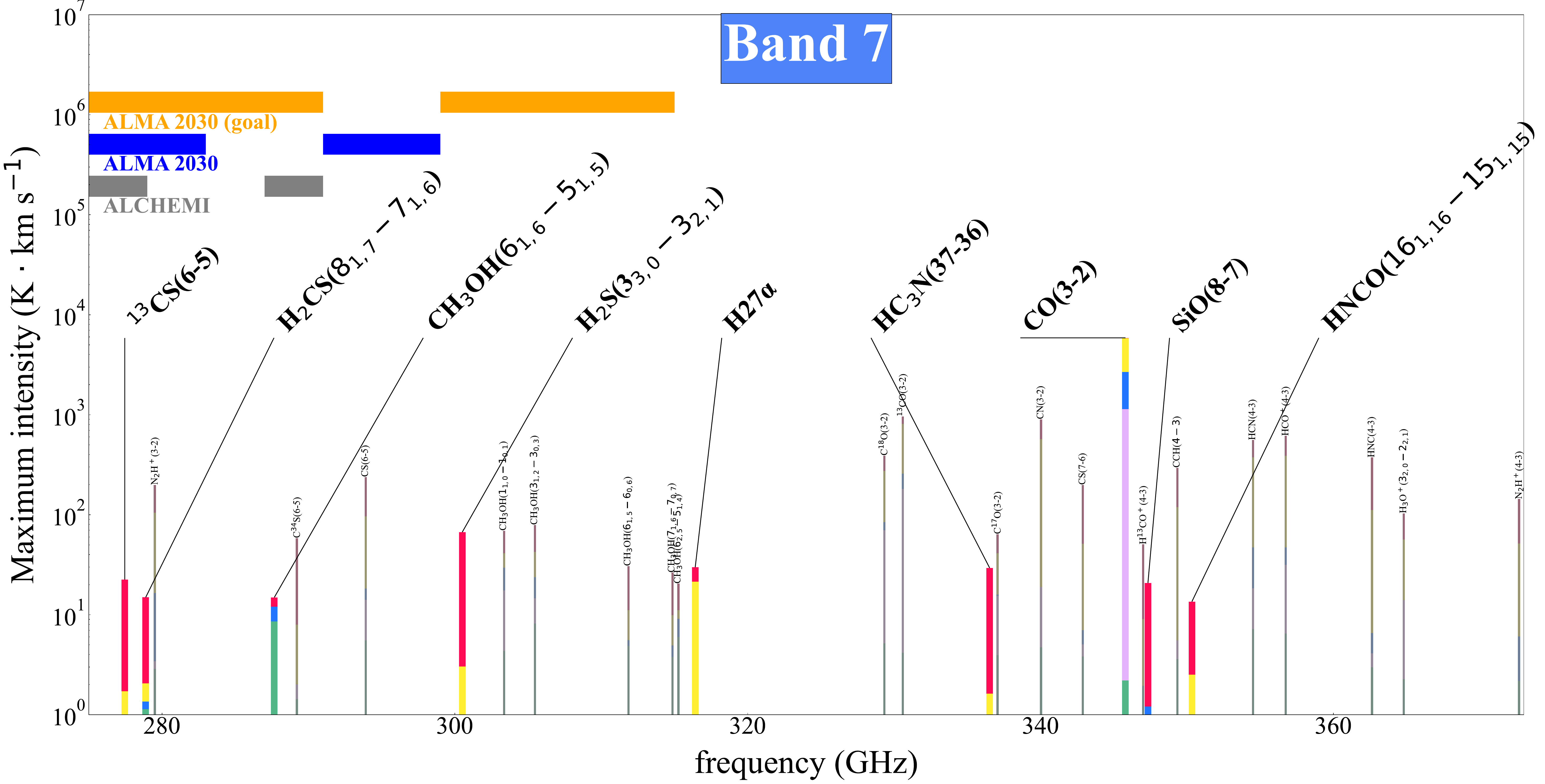}
    \end{center}
\caption{Same as figure \ref{fig:Useful_tracers_band3}, but for Band 7.}
\label{fig:Useful_tracers_band7}
\end{figure*}

    In actual observations, it is important to examine the distribution of the transition lines or continuum emissions that well explain the physical phenomena related to star formation. In this section, we will discuss how the tracers extracted by NMF can be used. Having the knowledge of tracers has an important implication in the era of the ALMA2030 Wideband Sensitivity Upgrade (WSU) \citep{ALMAmemo621} and ngVLA. With the extended bandwidth after WSU, we can simultaneously observe a greater number of transitions in a single-frequency setting. It will be possible to study various physical phenomena with a small number of tunings including all the useful tracers.
    
    Figures \ref{fig:Useful_tracers_band3} to \ref{fig:Useful_tracers_band7} show the intensity of transition lines and the ratio of the NMF components for each band. In these transition lines, a tracer was defined as one with an NF fraction of 0.6 or greater for any of the components. 
    %We only discuss the NF1 and NF3--5, because there are no transition lines with NF2 greater than 0.6. The tracers are colored, the lines are thickened, and the text is made to stand out. The horizontal axis is frequency, the vertical axis is maximum intensity of pixels in each integrated intensity map, since there are no significant outliers in the pixels. At the top are the spectral widths that can be observed in one tuning at ALCHEMI (current ALMA), ALMA2030, and ALMA2030 (goal). 
    It can be seen that NF1 tracers are relatively abundant in bands 4 and 5, NF2 in 3, 5, and 6, NF3 in bands 3 and 4, NF4 in bands 3 and 7, and NF5 in bands 7. If we consider extracting all tracers using current ALMA without considering the intensity of the transition lines, we could obtain the information on the transition lines and integrated intensity maps for all NF tracers by tuning twice in band 3, with 84\,GHz at the lower end of the LSB (lower sideband). Specifically, we may use the following transition lines as tracers: HC$_5$N(34-33) is tracer of NF1, CCH(1-0), HCN(1-0), HCO$^+$(1-0), and HOC$^+$(1-0) are tracers of NF2, HOCO$^+$($4_{0,4}-3_{0,3}$), HNCO($4_{0,4}-3_{0,3}$), OCS(8-7), and CH$_3$OH($2_{1,1}-1_{1,0}$) are tracers of NF3, H52$\beta$, H58$\gamma$, and H40$\alpha$ are tracers of NF4, and HC$_3$N,v$_7$=1(11$_1$-10$_{-1}$) are tracer of NF5. 
    
    In practice, however, it is desirable to obtain emission lines with as high an intensity as possible from the viewpoint of signal-to-noise ratio, etc. Therefore, it is also important to obtain transition lines with a reasonably high intensity. Perhaps we should observe transition lines like HNCO($9_{0,9}-8_{0,8}$) (Band 5, tracer of NF1) and NF5 tracers in Band 6 and 7 which have relatively high intensity of $\sim$ 100$\mathrm{\,K \cdot km \, s^{-1}}$. To improve the accuracy of the observations and data analysis, it will be necessary to observe these lines as well.

% Emission lines that have not been considered as tracers in the past are also considered to be tracers. Among them, those with particularly high NF values could be used as new tracers. Among these, the following are considered particularly useful. The high energy value of HOCO$^+$ (shock) has high percentage of NF3 value. This may be useful as a tracer for new shocks, not only because of the high percentage of NF3, but also because the value itself is relatively high. CCH is among those with a high percentage of NF4 (starburst). Like HOCO$^+$, this one has not only a high percentage of NF4, but also a relatively large value itself, making it a useful tracer of new star formation. Some with particularly high percentages of NF5 (early star formation) are CCS (especially at high levels). These molecules may be tracers of new early star formation.

\section{Summary}\label{sec:summary}
In this study, we attempted to extract the molecular gas features that dominate star formation and the molecules that trace them from the images of 148 different transition lines and continua at two frequency bands acquired by the ALCHEMI survey. We used a dimensionality reduction method called non-negative matrix factorization (NMF) to deal with the large-size data effectively. We found the following things. 
\begin{itemize}
    \item We extract five components related to star formation. These are the distribution of i) the middle-excited shock tracer, ii) low-density gas, iii) shock (cloud-cloud collision), iv) starburst, and v) young star formation, which we name NFs1--5.
    \item The characteristic transition lines or continuum emissions for each component is
        \begin{enumerate}
            \item Intermediate-excitation transitions of HOCO$^+$, HNCO and CH$_3$OH (NF1 = shock tracers with medium excitation)
            \item CO, CO isotopologues and HOC$^+$ (NF2 = low-density gas)
            \item Low-excitation transitions of HOCO$^+$, HNCO and CH$_3$OH (NF3 = shocks)
            \item Radio recombination lines (RRLs), CN, and CCH (NF4 = starbursts)
            \item High-excitation transitions of HC$_3$N, CCS, SO$_2$, and so on (NF5 = young star formation)
        \end{enumerate}
    \item Compared to the previous study by \citet{Harada2024}, it is easier to interpret the physical components of NMF and it is also a better method for independently extracting the diffuse components involved in outflows powered by star formation.
    \item NF1, 3, and 5 also represent the energy levels of the transition lines. As the percentage of NF5 in the transition lines or continuum increases, their energy state increases. Conversely, as the percentage of NF3 increases, the energy state of the transition lines decreases. NF1 represents middle energy state emission lines.
    \item NMF is superior in that it can extract localized and dilute structures without being affected by overall molecular distribution.
    \item The NF tracers, which have the high NF$x$ fractions, are gathered in Band 3. Therefore, we can extract the structure involved in star formation by observations in about 85\,GHz to 100\,GHz region of Band 3.
\end{itemize}

\bigskip
\begin{ack}
We are grateful to the summer student program at the National Astronomical Observatory of Japan / Astronomical Science Program of SOKENDAI, where this research initially took place. This paper makes use of the following ALMA data: ADS/JAO.ALMA\#2017.1.00161.L, ADS/JAO.ALMA\#2018.1.00162.S. ALMA is a partnership of ESO (representing its member states), NSF (USA) and NINS (Japan), together with NRC (Canada), MOST and ASIAA (Taiwan), and KASI (Republic of Korea), in cooperation with the Republic of Chile. The National Radio Astronomy Observatory is a facility of the National Science Foundation operated under cooperative agreement by Associated Universities, Inc. The Joint ALMA Observatory is operated by ESO, AUI/NRAO and NAOJ. Data analysis was in part carried out on the Multi-wavelength Data Analysis System operated by the Astronomy Data Center (ADC), National Astronomical Observatory of Japan. N.H. acknowledges support from JSPS KAKENHI grant No. JP21K03634.
V.M.R. and L.C. acknowledge support from the grant No. PID2022-136814NB-I00 by the Spanish Ministry of Science, Innovation and Universities/State Agency of Research MICIU/AEI/10.13039/501100011033 and by ERDF, UE. V.M.R also acknowledges support from project number RYC2020-029387-I funded by MICIU/AEI/10.13039/501100011033 and by "ESF, Investing in your future", and from the Consejo Superior de Investigaciones Cient{\'i}ficas (CSIC) and the Centro de Astrobiolog{\'i}a (CAB) through the project 20225AT015 (Proyectos intramurales especiales del CSIC), and from the grant CNS2023-144464 funded by MICIU/AEI/10.13039/501100011033 and by “European Union NextGenerationEU/PRTR”. S.V. acknowledges support by the European Research council (ERC) Advanced Grant MOPPEX 833460. S.A. gratefully acknowledges support from the ERC Advanced Grant 789410.
\end{ack}

\appendix
\section{Derivation method of base and coefficient matrices in scikit-learn}\label{appx: NMF_params}
To obtain the base matrix \textsf{B} and coefficient matrix \textsf{C}, the algorithm in \texttt{scikit-learn} minimizes the following function called the objective function\footnote{$\langle$ https://scikit-learn.org/stable/modules/generated/sklearn.decomposition.NMF.html $\rangle$}.
\begin{eqnarray}
    \begin{aligned}
        L(\textsf{B}, \textsf{C}) &= 0.5 \times \|\textsf{E} - \textsf{B} \textsf{C} \|^2_\mathrm{loss} \\
        &+ \alpha_W \times l_{1r} \times N \times \|vec(\textsf{B} \|_1  \\
        &+ \alpha_H \times l_\mathrm{1r} \times L \times \|vec(\textsf{C})\|_1 \\
        &+ 0.5 \times \alpha_W \times (1 - l_{1r}) \times N \times \| \textsf{B}\|_{Fro}^2 \\
        &+ 0.5 \times \alpha_H \times (1 - l_{1r}) \times L \times \| \textsf{C}\|_{Fro}^2
    \end{aligned}
\end{eqnarray}
where $\| A\|_\mathrm{loss}$ is a loss function, $\| A\|_\mathrm{Fro}$ is the Frobenius norm derived by \[\|A\|_{Fro}^2=\sum_{i,j}(A_{i,j})^2,\] $\|vec(A)\|_1$ is an elementwise L$^1$ norm \[ \|vec(A)\|_1 = \sum_{i, j}  abs(\textsf{B}_{i,j}),\] and $abs(a)$ is the absolute value of $a$. The indices $N$ and $L$ are the same as the ones defined in Section \ref{sec:method}. The loss function is commonly chosen to be the Frobenius norm, which we use here. Variables $\alpha_\mathrm{W}$, $\alpha_\mathrm{H}$, and $l_\mathrm{{1r}}$ corresponds to parameters $\mathrm{alpha\_W}$, $\mathrm{alpha\_H}$, and $\mathrm{l1\_ratio}$ used in \texttt{scikit-learn}, respectively. The parameters used for this study and their meanings are shown in table \ref{table:parameters}. Note that, for the parameters we used, the objective function simply becomes the Frobenius norm of \textsf{E} $-$ \textsf{BC}. 

For a given set of solutions for \textsf{B} and \textsf{C}, one can create multiple combinations of matrices \textsf{B'} and \textsf{C'} that satisfy the condition \textsf{BC} = \textsf{B'C'}. For example, imagine a $5\times5$ diagonal  matrix \textsf{A} with diagonal entries of $a_1$, $a_2$, $a_3$, $a_4$, and $a_5$. Its inverse matrix \textsf{A$^{-1}$} will be a diagonal matrix with diagonal entries of $1/a_1$, $1/a_2$, $1/a_3$, $1/a_4$, and $1/a_5$. By taking \textsf{B'} = \textsf{BA$^{-1}$} and \textsf{C'} = \textsf{AC}, it gives \textsf{B'C'} = \textsf{BC}. A set of matrices \textsf{B'} and \textsf{C'} is also a solution of NMF. In other words, one can scale NF maps arbitrarily. To avoid NF components from being unevenly reflected in coefficients, we scale the NF maps so that their L$^1$ norms ($\sum_{l=1}^L abs(b'_{l1})$, $\sum_{l=1}^L abs(b'_{l2})$, ..., $\sum_{l=1}^L abs(b'_{l5})$) become 1. Using the notation above, the L$^1$ norms of the columns of \textsf{B} are the scaling factors $a_1$, $a_2$, etc. Subsequently, we calculate the corresponding coefficient matrix.

After scaling the base matrices, coefficients directly reflect how much flux each NF contributes to the total flux of input images. When we reconstruct the input image, the $l$-th pixel of the $n$-th image will have the flux \(\sum_{m=1}^M b_{lm}c_{mn}.\) If we sum the flux from all the pixels, \(\sum_{l=1}^L\sum_{m=1}^M b_{lm}c_{mn} = \sum_{m=1}^M c_{mn}\) because the fact that $b_{lm}>0$ leads to \(\sum_{l=1}^L b_{lm} = 1\). 

Depending on the choice of initial values of matrices, the objective function may only converge to a local minimum. To make sure that it is unlikely the case, we confirmed that changing the initialization methods does not change our results when using standard methods of initialization in \texttt{scikit-learn}, nndsvda, nndsvd, nndsvdar, and random. 

\begin{table*}[ht]
    \tbl{The parameters of NMF using scikit-learn.}{%}
    \centering
    \begin{tabular}{ccl}
    \hline
    Parameters of NMF & Value in this study & \multicolumn{1}{c}{Meaning of the parameter} \\ \hline
    n\_components & 5 & The number of components \\
    init & nndsvda & Initialization method \footnotemark[$*$] \\
    solver & cd & the optimization algorithm\footnotemark[$\dag$] \\
    beta\_loss & frobenius & The distance between $\bm{\textsf{E}}$ and $\bm{\textsf{B}} \bm{\textsf{C}}$ \\
    \multirow{2}{*}{tol} & \multirow{2}{*}{0.0001} & Once the difference of the loss between two \\
    & & consecutive iteration falls within 'tol', the calculation stops. \\
    max\_iter & 1000 & Maximum number of times until the iteration stops. \\
    random\_state & None & Whether to fix the random number seed. \\
    alpha\_W & 0.0 & The one of the parameters in the loss function. \\
    alpha\_H & 0.0 & The one of the parameters in the loss function. \\
    \multirow{2}{*}{l1\_ratio} & \multirow{2}{*}{0.0} & The regularization parameter to avoid over-fitting, \\
    & & and the one of the parameters in the loss function. \\
    verbose & 0 & Whether display the situation of iteration progress. \\
    shuffle & None & Whether randomize the order of coordinates. \\ \hline
    \end{tabular}}
    \begin{tabnote}
        \footnotemark[$*$] \citet{Boutsidis2007}
        \footnotemark[$\dag$] \citet{Cichocki2009}
    \end{tabnote}
    \label{table:parameters}
\end{table*}

\section{The error before and after applying NMF}
\begin{figure}[ht]
    \begin{center}
        \includegraphics[keepaspectratio, width=\linewidth]{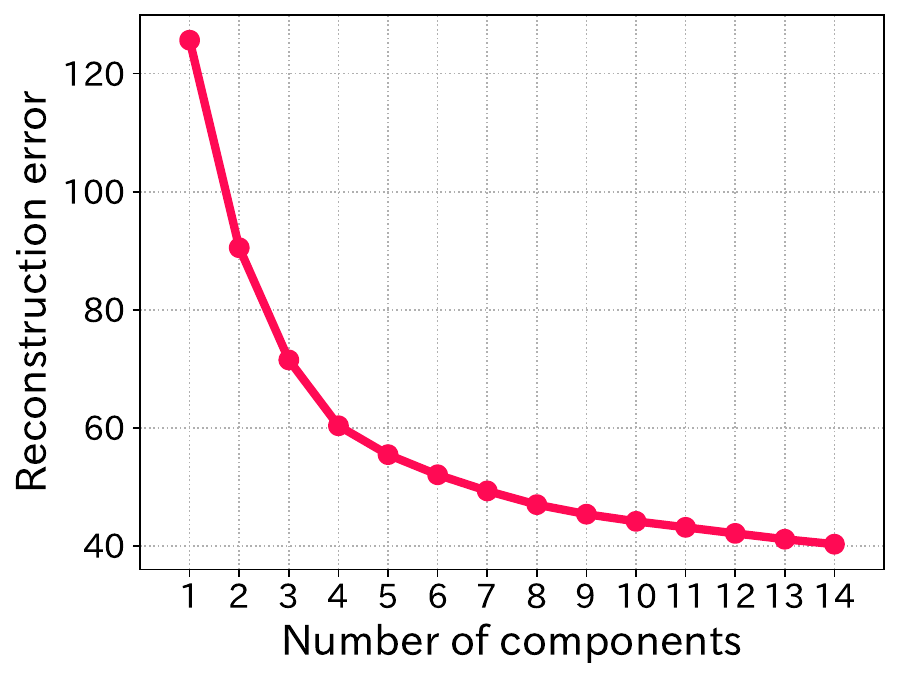}
    \end{center}
\caption{The reconstruction error for each NMF component represented as red dots, and they are connected by the lines of same color. As the number of NMF component increases, the error decreases.}
\label{fig:error}
\end{figure}

The error between $\bm{\textsf{E}}$ and reconstructed $\bm{\textsf{BC}}$ (reconstruction error) is expressed by the Frobenius norm of \textsf{E} $-$ \textsf{BC}. We show the variation of the reconstruction error with number of components in figure \ref{fig:error}. The reconstruction error decreases steeply with the number of components for $M<4$, but it plateaus for larger $M$.

\section{The case of another number of components}
\begin{figure*}[ht]
    \begin{center}        \includegraphics[keepaspectratio, width=\linewidth]{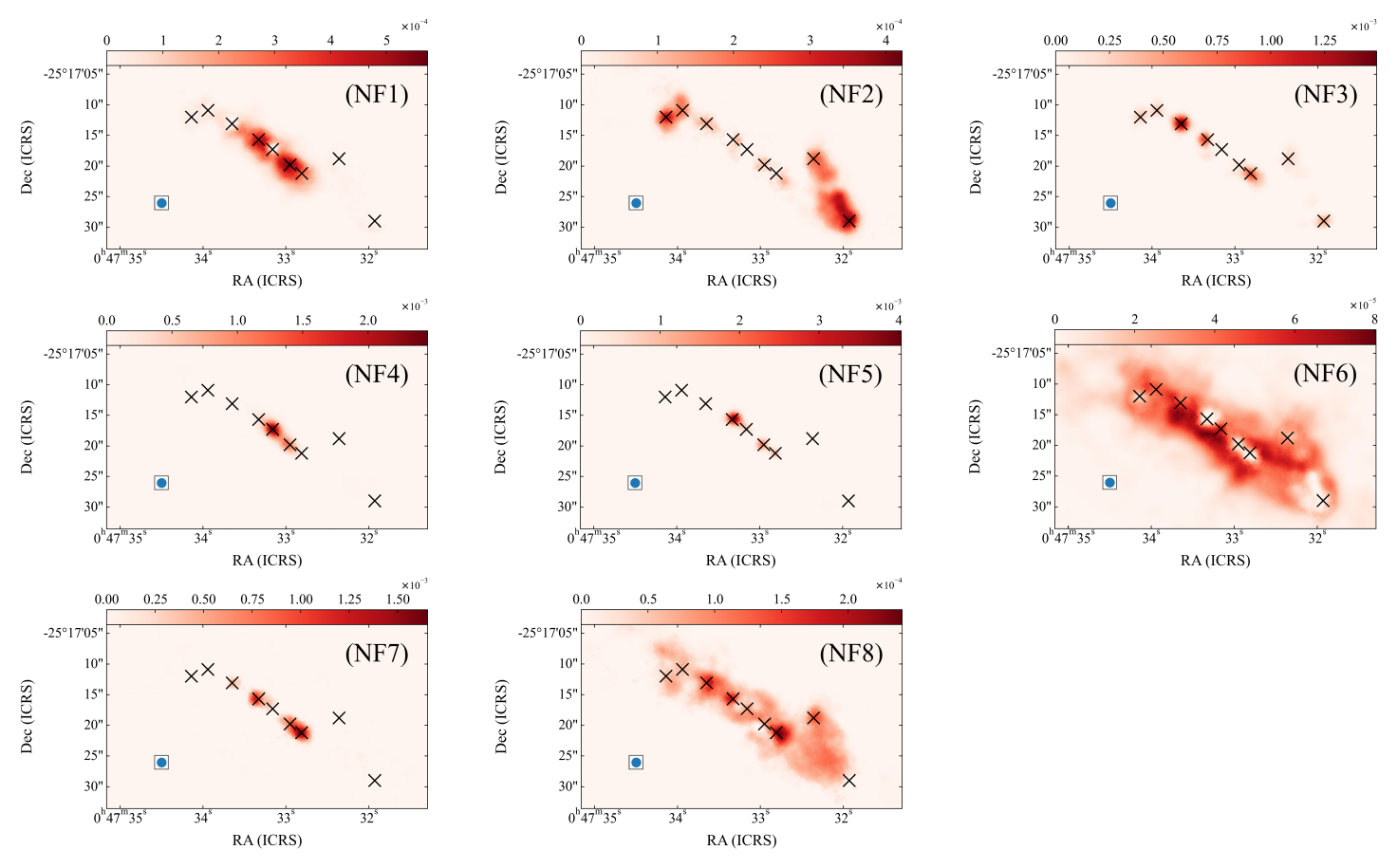}
    \end{center}
\caption{Maps of NF base matrices when the number of components is 8.}
\label{fig:comp7_all}
\end{figure*}

The NF maps in the case where we increase the number of components from five to eight are shown in figure \ref{fig:comp7_all}. As the number of components increases, more components contain structures that are too localized or noisy. We conclude that it defeats the purpose of dimensionality reduction to use the number of components that is larger than seven.

% To solve this problem, we defined a measure of how far apart the two matrices are. While there are many different measures of this metric, we have used the Euclidean distance. the Euclidean distance is defined below.
% \begin{align}
%     \| \textsf{E - BC} \|^2 = \sum_{i, j} [\textsf{(E - BC)}_{ij}]^2
% \end{align}

% In this study, we assumed termination if $\| \textsf{E - BC} \| < 10^{-4}$

\bibliography{pasj}

\begin{thebibliography}{}
\expandafter\ifx\csname natexlab\endcsname\relax\def\natexlab#1{#1}\fi
\providecommand{\url}[1]{\href{#1}{#1}}
\providecommand{\dodoi}[1]{doi:~\href{http://doi.org/#1}{\nolinkurl{#1}}}
\providecommand{\doeprint}[1]{\href{http://ascl.net/#1}{\nolinkurl{http://ascl.net/#1}}}
\providecommand{\doarXiv}[1]{\href{https://arxiv.org/abs/#1}{\nolinkurl{https://arxiv.org/abs/#1}}}

\bibitem[{{Ando} {et~al.}(2017){Ando}, {Nakanishi}, {Kohno}, {Izumi}, {Mart{\'\i}n}, {Harada}, {Takano}, {Kuno}, {Nakai}, {Sugai}, {Sorai}, {Tosaki}, {Matsubayashi}, {Nakajima}, {Nishimura}, \& {Tamura}}]{Ando2017}
{Ando}, R., {Nakanishi}, K., {Kohno}, K., {et~al.} 2017, \apj, 849, 81

\bibitem[{{Bao} {et~al.}(2024){Bao}, {Harada}, {Kohno}, {Yoshimura}, {Egusa}, {Nishimura}, {Tanaka}, {Nakanishi}, {Mart{\'\i}n}, {Mangum}, {Sakamoto}, {Muller}, {Bouvier}, {Colzi}, {Emig}, {Meier}, {Henkel}, {Humire}, {Huang}, {Rivilla}, {van der Werf}, \& {Viti}}]{Bao2024}
{Bao}, M., {Harada}, N., {Kohno}, K., {et~al.} 2024, arXiv e-prints, arXiv:2404.04791

\bibitem[{{Barrientos} {et~al.}(2021){Barrientos}, {Holdship}, {Solar}, {Mart{\'\i}n}, {Rivilla}, {Viti}, {Mangum}, {Harada}, {Sakamoto}, {Muller}, {Tanaka}, {Yoshimura}, {Nakanishi}, {Herrero-Illana}, {M{\"u}hle}, {Aladro}, {Aalto}, {Henkel}, \& {Humire}}]{Barrientos2021}
{Barrientos}, A., {Holdship}, J., {Solar}, M., {et~al.} 2021, Experimental Astronomy, 52, 157

\bibitem[{{Behrens} {et~al.}(2022){Behrens}, {Mangum}, {Holdship}, {Viti}, {Harada}, {Mart{\'\i}n}, {Sakamoto}, {Muller}, {Tanaka}, {Nakanishi}, {Herrero-Illana}, {Yoshimura}, {Aladro}, {Colzi}, {Emig}, {Henkel}, {Huang}, {Humire}, {Meier}, {Rivilla}, {van der Werf}, \& {Alma Comprehensive High-Resolution Extragalactic Molecular Inventory (Alchemi) Collaboration}}]{Behrens2022}
{Behrens}, E., {Mangum}, J.~G., {Holdship}, J., {et~al.} 2022, \apj, 939, 119

\bibitem[{{Bendo} {et~al.}(2015){Bendo}, {Beswick}, {D'Cruze}, {Dickinson}, {Fuller}, \& {Muxlow}}]{Bendo2013}
{Bendo}, G.~J., {Beswick}, R.~J., {D'Cruze}, M.~J., {et~al.} 2015, \mnras, 450, L80

\bibitem[{{Bolatto} {et~al.}(2013){Bolatto}, {Warren}, {Leroy}, {Walter}, {Veilleux}, {Ostriker}, {Ott}, {Zwaan}, {Fisher}, {Weiss}, {Rosolowsky}, \& {Hodge}}]{Bolatto2013}
{Bolatto}, A.~D., {Warren}, S.~R., {Leroy}, A.~K., {et~al.} 2013, \nat, 499, 450

\bibitem[{Boutsidis \& Gallopoulos(2008)}]{Boutsidis2007}
Boutsidis, C., \& Gallopoulos, E. 2008, Pattern Recognition, 41, 1350

\bibitem[{{Bouvier} {et~al.}(2024){Bouvier}, {Viti}, {Behrens}, {Butterworth}, {Huang}, {Mangum}, {Harada}, {Mart{\'\i}n}, {Rivilla}, {Muller}, {Sakamoto}, {Yoshimura}, {Tanaka}, {Nakanishi}, {Herrero-Illana}, {Colzi}, {Gorski}, {Henkel}, {Humire}, {Meier}, {van der Werf}, \& {Yan}}]{2024arXiv240508408B}
{Bouvier}, M., {Viti}, S., {Behrens}, E., {et~al.} 2024, arXiv e-prints, arXiv:2405.08408

\bibitem[{{Brunthaler} {et~al.}(2009){Brunthaler}, {Castangia}, {Tarchi}, {Henkel}, {Reid}, {Falcke}, \& {Menten}}]{Brunthaler2009}
{Brunthaler}, A., {Castangia}, P., {Tarchi}, A., {et~al.} 2009, \aap, 497, 103

\bibitem[{{Butterworth} {et~al.}(2024){Butterworth}, {Viti}, {Van der Werf}, {Mangum}, {Mart{\'\i}n}, {Harada}, {Emig}, {Muller}, {Sakamoto}, {Yoshimura}, {Tanaka}, {Herrero-Illana}, {Colzi}, {Rivilla}, {Huang}, {Bouvier}, {Behrens}, {Henkel}, {Yan}, {Meier}, \& {Zhou}}]{2024A&A...686A..31B}
{Butterworth}, J., {Viti}, S., {Van der Werf}, P.~P., {et~al.} 2024, \aap, 686, A31

\bibitem[{Carpenter {et~al.}(2022)Carpenter, Brogan, Iono, \& Mroczkowski}]{ALMAmemo621}
Carpenter, J.~M., Brogan, C.~L., Iono, D., \& Mroczkowski, T. 2022, ALMA Memo 621, arXiv:2211.00195

\bibitem[{{Chen} {et~al.}(2018){Chen}, {Ellingsen}, {Shen}, {McCarthy}, {Zhong}, \& {Deng}}]{Chen2018}
{Chen}, X., {Ellingsen}, S.~P., {Shen}, Z.-Q., {et~al.} 2018, \apjl, 856, L35

\bibitem[{Cichock \& Phan(2009)}]{Cichocki2009}
Cichock, A., \& Phan, A.-H. 2009, IEICE Transactions on Fundamentals of Electronics, Communications and Computer Sciences, E92.A, 708

\bibitem[{{de Mijolla} {et~al.}(2024){de Mijolla}, {Holdship}, {Viti}, \& {Heyl}}]{Mijolla2024}
{de Mijolla}, D., {Holdship}, J., {Viti}, S., \& {Heyl}, J. 2024, \apj, 961, 225

\bibitem[{{Efstathiou} \& {Fall}(1984)}]{1984MNRAS.206..453E}
{Efstathiou}, G., \& {Fall}, S.~M. 1984, \mnras, 206, 453

\bibitem[{{Ellingsen} {et~al.}(2017){Ellingsen}, {Chen}, {Breen}, \& {Qiao}}]{Ellingsen2017}
{Ellingsen}, S.~P., {Chen}, X., {Breen}, S.~L., \& {Qiao}, H.~H. 2017, \mnras, 472, 604

\bibitem[{{Gao} \& {Solomon}(2004)}]{Gao2004}
{Gao}, Y., \& {Solomon}, P.~M. 2004, \apj, 606, 271

\bibitem[{{Gorski} {et~al.}(2019){Gorski}, {Ott}, {Rand}, {Meier}, {Momjian}, {Schinnerer}, \& {Ellingsen}}]{Gorski2019}
{Gorski}, M.~D., {Ott}, J., {Rand}, R., {et~al.} 2019, \mnras, 483, 5434

\bibitem[{{Gratier} {et~al.}(2017){Gratier}, {Bron}, {Gerin}, {Pety}, {Guzman}, {Orkisz}, {Bardeau}, {Goicoechea}, {Le Petit}, {Liszt}, {{\"O}berg}, {Peretto}, {Roueff}, {Sievers}, \& {Tremblin}}]{Gratier2017}
{Gratier}, P., {Bron}, E., {Gerin}, M., {et~al.} 2017, \aap, 599, A100

\bibitem[{{Grudi{\'c}} {et~al.}(2018){Grudi{\'c}}, {Hopkins}, {Faucher-Gigu{\`e}re}, {Quataert}, {Murray}, \& {Kere{\v{s}}}}]{Grudic2018}
{Grudi{\'c}}, M.~Y., {Hopkins}, P.~F., {Faucher-Gigu{\`e}re}, C.-A., {et~al.} 2018, \mnras, 475, 3511

\bibitem[{{Haasler} {et~al.}(2022){Haasler}, {Rivilla}, {Mart{\'\i}n}, {Holdship}, {Viti}, {Harada}, {Mangum}, {Sakamoto}, {Muller}, {Tanaka}, {Yoshimura}, {Nakanishi}, {Colzi}, {Hunt}, {Emig}, {Aladro}, {Humire}, {Henkel}, \& {van der Werf}}]{Haasler2022}
{Haasler}, D., {Rivilla}, V.~M., {Mart{\'\i}n}, S., {et~al.} 2022, \aap, 659, A158

\bibitem[{{Harada} {et~al.}(2021){Harada}, {Mart{\'\i}n}, {Mangum}, {Sakamoto}, {Muller}, {Tanaka}, {Nakanishi}, {Herrero-Illana}, {Yoshimura}, {M{\"u}hle}, {Aladro}, {Colzi}, {Rivilla}, {Aalto}, {Behrens}, {Henkel}, {Holdship}, {Humire}, {Meier}, {Nishimura}, {van der Werf}, \& {Viti}}]{Harada2021}
{Harada}, N., {Mart{\'\i}n}, S., {Mangum}, J.~G., {et~al.} 2021, \apj, 923, 24

\bibitem[{{Harada} {et~al.}(2022){Harada}, {Mart{\'\i}n}, {Mangum}, {Sakamoto}, {Muller}, {Rivilla}, {Henkel}, {Meier}, {Colzi}, {Yamagishi}, {Tanaka}, {Nakanishi}, {Herrero-Illana}, {Yoshimura}, {Humire}, {Aladro}, {van der Werf}, \& {Emig}}]{Harada2022}
---. 2022, \apj, 938, 80

\bibitem[{{Harada} {et~al.}(2024){Harada}, {Meier}, {Mart{\'\i}n}, {Muller}, {Sakamoto}, {Saito}, {Gorski}, {Henkel}, {Tanaka}, {Mangum}, {Aalto}, {Aladro}, {Bouvier}, {Colzi}, {Emig}, {Herrero-Illana}, {Huang}, {Kohno}, {K{\"o}nig}, {Nakanishi}, {Nishimura}, {Takano}, {Rivilla}, {Viti}, {Watanabe}, {van der Werf}, \& {Yoshimura}}]{Harada2024}
{Harada}, N., {Meier}, D.~S., {Mart{\'\i}n}, S., {et~al.} 2024, arXiv e-prints, arXiv:2401.02578

\bibitem[{{Holdship} {et~al.}(2021){Holdship}, {Viti}, {Mart{\'\i}n}, {Harada}, {Mangum}, {Sakamoto}, {Muller}, {Tanaka}, {Yoshimura}, {Nakanishi}, {Herrero-Illana}, {M{\"u}hle}, {Aladro}, {Colzi}, {Emig}, {Garc{\'\i}a-Burillo}, {Henkel}, {Humire}, {Meier}, {Rivilla}, \& {van der Werf}}]{Holdship2021}
{Holdship}, J., {Viti}, S., {Mart{\'\i}n}, S., {et~al.} 2021, \aap, 654, A55

\bibitem[{{Holdship} {et~al.}(2022){Holdship}, {Mangum}, {Viti}, {Behrens}, {Harada}, {Mart{\'\i}n}, {Sakamoto}, {Muller}, {Tanaka}, {Nakanishi}, {Herrero-Illana}, {Yoshimura}, {Aladro}, {Colzi}, {Emig}, {Henkel}, {Nishimura}, {Rivilla}, {van der Werf}, \& {Alma Comprehensive High-Resolution Extragalactic Molecular Inventory (Alchemi) Collaboration}}]{Holdship2022}
{Holdship}, J., {Mangum}, J.~G., {Viti}, S., {et~al.} 2022, \apj, 931, 89

\bibitem[{{Huang} {et~al.}(2023){Huang}, {Viti}, {Holdship}, {Mangum}, {Mart{\'\i}n}, {Harada}, {Muller}, {Sakamoto}, {Tanaka}, {Yoshimura}, {Herrero-Illana}, {Meier}, {Behrens}, {van der Werf}, {Henkel}, {Garc{\'\i}a-Burillo}, {Rivilla}, {Emig}, {Colzi}, {Humire}, {Aladro}, \& {Bouvier}}]{Huang2023}
{Huang}, K.~Y., {Viti}, S., {Holdship}, J., {et~al.} 2023, \aap, 675, A151

\bibitem[{{Humire} {et~al.}(2022){Humire}, {Henkel}, {Hern{\'a}ndez-G{\'o}mez}, {Mart{\'\i}n}, {Mangum}, {Harada}, {Muller}, {Sakamoto}, {Tanaka}, {Yoshimura}, {Nakanishi}, {M{\"u}hle}, {Herrero-Illana}, {Meier}, {Caux}, {Aladro}, {Mauersberger}, {Viti}, {Colzi}, {Rivilla}, {Gorski}, {Menten}, {Huang}, {Aalto}, {van der Werf}, \& {Emig}}]{Humire2022}
{Humire}, P.~K., {Henkel}, C., {Hern{\'a}ndez-G{\'o}mez}, A., {et~al.} 2022, \aap, 663, A33

\bibitem[{{Hurley} {et~al.}(2014){Hurley}, {Oliver}, {Farrah}, {Lebouteiller}, \& {Spoon}}]{2014MNRAS.437..241H}
{Hurley}, P.~D., {Oliver}, S., {Farrah}, D., {Lebouteiller}, V., \& {Spoon}, H.~W.~W. 2014, \mnras, 437, 241

\bibitem[{Jolliffe(2002)}]{jolliffe2002}
Jolliffe, I. 2002, {Principal Component Analysis}, Springer Series in Statistics (Springer)

\bibitem[{{Kennicutt} \& {Evans}(2012)}]{Kennicutt2012}
{Kennicutt}, R.~C., \& {Evans}, N.~J. 2012, \araa, 50, 531

\bibitem[{{Krieger} {et~al.}(2019){Krieger}, {Bolatto}, {Walter}, {Leroy}, {Zschaechner}, {Meier}, {Ott}, {Weiss}, {Mills}, {Levy}, {Veilleux}, \& {Gorski}}]{Krieger2019}
{Krieger}, N., {Bolatto}, A.~D., {Walter}, F., {et~al.} 2019, \apj, 881, 43

\bibitem[{{Lada} {et~al.}(2010){Lada}, {Lombardi}, \& {Alves}}]{Lada2010}
{Lada}, C.~J., {Lombardi}, M., \& {Alves}, J.~F. 2010, \apj, 724, 687

\bibitem[{{Lee} \& {Seung}(1999)}]{Lee1999}
{Lee}, D.~D., \& {Seung}, H.~S. 1999, \nat, 401, 788

\bibitem[{{Lee} \& {Seung}(2001)}]{Lee2001}
---. 2001, Adv. Neural Inf. Proc. Syst., 13, 556

\bibitem[{{Leroy} {et~al.}(2015){Leroy}, {Bolatto}, {Ostriker}, {Rosolowsky}, {Walter}, {Warren}, {Donovan Meyer}, {Hodge}, {Meier}, {Ott}, {Sandstrom}, {Schruba}, {Veilleux}, \& {Zwaan}}]{Leroy2015}
{Leroy}, A.~K., {Bolatto}, A.~D., {Ostriker}, E.~C., {et~al.} 2015, \apj, 801, 25

\bibitem[{{Leroy} {et~al.}(2018){Leroy}, {Bolatto}, {Ostriker}, {Walter}, {Gorski}, {Ginsburg}, {Krieger}, {Levy}, {Meier}, {Mills}, {Ott}, {Rosolowsky}, {Thompson}, {Veilleux}, \& {Zschaechner}}]{Leroy2018}
---. 2018, \apj, 869, 126

\bibitem[{{Lupton} {et~al.}(2004){Lupton}, {Blanton}, {Fekete}, {Hogg}, {O'Mullane}, {Szalay}, \& {Wherry}}]{Lupton2004}
{Lupton}, R., {Blanton}, M.~R., {Fekete}, G., {et~al.} 2004, \pasp, 116, 133

\bibitem[{{Madau} \& {Dickinson}(2014)}]{Madau2014}
{Madau}, P., \& {Dickinson}, M. 2014, \araa, 52, 415

\bibitem[{{Mart{\'\i}n} {et~al.}(2021){Mart{\'\i}n}, {Mangum}, {Harada}, {Costagliola}, {Sakamoto}, {Muller}, {Aladro}, {Tanaka}, {Yoshimura}, {Nakanishi}, {Herrero-Illana}, {M{\"u}hle}, {Aalto}, {Behrens}, {Colzi}, {Emig}, {Fuller}, {Garc{\'\i}a-Burillo}, {Greve}, {Henkel}, {Holdship}, {Humire}, {Hunt}, {Izumi}, {Kohno}, {K{\"o}nig}, {Meier}, {Nakajima}, {Nishimura}, {Padovani}, {Rivilla}, {Takano}, {van der Werf}, {Viti}, \& {Yan}}]{Martin2021}
{Mart{\'\i}n}, S., {Mangum}, J.~G., {Harada}, N., {et~al.} 2021, \aap, 656, A46

\bibitem[{{Meier} \& {Turner}(2005)}]{Meier2005}
{Meier}, D.~S., \& {Turner}, J.~L. 2005, \apj, 618, 259

\bibitem[{{Meier} {et~al.}(2015){Meier}, {Walter}, {Bolatto}, {Leroy}, {Ott}, {Rosolowsky}, {Veilleux}, {Warren}, {Wei{\ss}}, {Zwaan}, \& {Zschaechner}}]{Meier2015}
{Meier}, D.~S., {Walter}, F., {Bolatto}, A.~D., {et~al.} 2015, \apj, 801, 63

\bibitem[{{M{\"u}ller-S{\'a}nchez} {et~al.}(2010){M{\"u}ller-S{\'a}nchez}, {Gonz{\'a}lez-Mart{\'\i}n}, {Fern{\'a}ndez-Ontiveros}, {Acosta-Pulido}, \& {Prieto}}]{Muller2010}
{M{\"u}ller-S{\'a}nchez}, F., {Gonz{\'a}lez-Mart{\'\i}n}, O., {Fern{\'a}ndez-Ontiveros}, J.~A., {Acosta-Pulido}, J.~A., \& {Prieto}, M.~A. 2010, \apj, 716, 1166

\bibitem[{Pedregosa {et~al.}(2011)Pedregosa, Varoquaux, Gramfort, Michel, Thirion, Grisel, Blondel, Prettenhofer, Weiss, Dubourg, Vanderplas, Passos, Cournapeau, Brucher, Perrot, \& Duchesnay}]{scikit-learn}
Pedregosa, F., Varoquaux, G., Gramfort, A., {et~al.} 2011, Journal of Machine Learning Research, 12, 2825

\bibitem[{{Rekola} {et~al.}(2005){Rekola}, {Richer}, {McCall}, {Valtonen}, {Kotilainen}, \& {Flynn}}]{Rekola2005}
{Rekola}, R., {Richer}, M.~G., {McCall}, M.~L., {et~al.} 2005, \mnras, 361, 330

\bibitem[{{Rico-Villas} {et~al.}(2020){Rico-Villas}, {Mart{\'\i}n-Pintado}, {Gonz{\'a}lez-Alfonso}, {Mart{\'\i}n}, \& {Rivilla}}]{Rico2020}
{Rico-Villas}, F., {Mart{\'\i}n-Pintado}, J., {Gonz{\'a}lez-Alfonso}, E., {Mart{\'\i}n}, S., \& {Rivilla}, V.~M. 2020, \mnras, 491, 4573

\bibitem[{{Saito} {et~al.}(2015){Saito}, {Iono}, {Yun}, {Ueda}, {Nakanishi}, {Sugai}, {Espada}, {Imanishi}, {Motohara}, {Hagiwara}, {Tateuchi}, {Lee}, \& {Kawabe}}]{Saito2015}
{Saito}, T., {Iono}, D., {Yun}, M.~S., {et~al.} 2015, \apj, 803, 60

\bibitem[{{Saito} {et~al.}(2022){Saito}, {Takano}, {Harada}, {Nakajima}, {Schinnerer}, {Liu}, {Taniguchi}, {Izumi}, {Watanabe}, {Bamba}, {Kohno}, {Nishimura}, {Stuber}, \& {Tosaki}}]{Saito2022}
{Saito}, T., {Takano}, S., {Harada}, N., {et~al.} 2022, \apj, 935, 155

\bibitem[{{Sakamoto} {et~al.}(2011){Sakamoto}, {Mao}, {Matsushita}, {Peck}, {Sawada}, \& {Wiedner}}]{Sakamoto2011}
{Sakamoto}, K., {Mao}, R.-Q., {Matsushita}, S., {et~al.} 2011, \apj, 735, 19

\bibitem[{{Sakamoto} {et~al.}(1999){Sakamoto}, {Okumura}, {Ishizuki}, \& {Scoville}}]{Sakamoto1999}
{Sakamoto}, K., {Okumura}, S.~K., {Ishizuki}, S., \& {Scoville}, N.~Z. 1999, \apj, 525, 691

\bibitem[{{Schilke} {et~al.}(1997){Schilke}, {Walmsley}, {Pineau des Forets}, \& {Flower}}]{Schilke1997}
{Schilke}, P., {Walmsley}, C.~M., {Pineau des Forets}, G., \& {Flower}, D.~R. 1997, \aap, 321, 293

\bibitem[{Spearman(1904)}]{Spearman1904}
Spearman, C. 1904, The American Journal of Psychology, 15, 72

\bibitem[{{Tanaka} {et~al.}(2024){Tanaka}, {Mangum}, {Viti}, {Mart{\'\i}n}, {Harada}, {Sakamoto}, {Muller}, {Yoshimura}, {Nakanishi}, {Herrero-Illana}, {Emig}, {M{\"u}hle}, {Kaneko}, {Tosaki}, {Behrens}, {Rivilla}, {Colzi}, {Nishimura}, {Humire}, {Bouvier}, {Huang}, {Butterworth}, {Meier}, \& {van der Werf}}]{Tanaka2024}
{Tanaka}, K., {Mangum}, J.~G., {Viti}, S., {et~al.} 2024, \apj, 961, 18

\bibitem[{{Traven} {et~al.}(2017){Traven}, {Matijevi{\v{c}}}, {Zwitter}, {{\v{Z}}erjal}, {Kos}, {Asplund}, {Bland-Hawthorn}, {Casey}, {De Silva}, {Freeman}, {Lin}, {Martell}, {Schlesinger}, {Sharma}, {Simpson}, {Zucker}, {Anguiano}, {Da Costa}, {Duong}, {Horner}, {Hyde}, {Kafle}, {Munari}, {Nataf}, {Navin}, {Reid}, \& {Ting}}]{2017ApJS..228...24T}
{Traven}, G., {Matijevi{\v{c}}}, G., {Zwitter}, T., {et~al.} 2017, \apjs, 228, 24

\bibitem[{{Turner} \& {Ho}(1985)}]{Turner1985}
{Turner}, J.~L., \& {Ho}, P.~T.~P. 1985, \apjl, 299, L77

\bibitem[{{Walter} {et~al.}(2017){Walter}, {Bolatto}, {Leroy}, {Veilleux}, {Warren}, {Hodge}, {Levy}, {Meier}, {Ostriker}, {Ott}, {Rosolowsky}, {Scoville}, {Weiss}, {Zschaechner}, \& {Zwaan}}]{Walter2017}
{Walter}, F., {Bolatto}, A.~D., {Leroy}, A.~K., {et~al.} 2017, \apj, 835, 265

\bibitem[{{Watanabe} {et~al.}(2016){Watanabe}, {Sakai}, {Sorai}, {Ueda}, \& {Yamamoto}}]{Watanabe2016}
{Watanabe}, Y., {Sakai}, N., {Sorai}, K., {Ueda}, J., \& {Yamamoto}, S. 2016, \apj, 819, 144

\bibitem[{{Weaver} {et~al.}(2002){Weaver}, {Heckman}, {Strickland}, \& {Dahlem}}]{Weaver2002}
{Weaver}, K.~A., {Heckman}, T.~M., {Strickland}, D.~K., \& {Dahlem}, M. 2002, \apjl, 576, L19

\bibitem[{{Weedman} {et~al.}(1981){Weedman}, {Feldman}, {Balzano}, {Ramsey}, {Sramek}, \& {Wuu}}]{Weedman1981}
{Weedman}, D.~W., {Feldman}, F.~R., {Balzano}, V.~A., {et~al.} 1981, \apj, 248, 105

\bibitem[{{Yamamoto}(2017)}]{Yamamoto2017}
{Yamamoto}, S. 2017, {Introduction to Astrochemistry: Chemical Evolution from Interstellar Clouds to Star and Planet Formation} (Springer Tokyo)

\end{thebibliography}


\begin{thebibliography}{}%%% references
\bibitem[Bolatto et~al.(2013)]{Bolatto2013} Bolatto, A.~D., et al. 2013, \nat, 499, 450
\bibitem[Harada et~al.(2023)]{Harada2023} Harada,~N., et al. 2023, 
\bibitem[Huang et~al.(2023)]{Huang2023} Huang, K.~-Y., et al. 2023, \aap, 675, A151
\bibi11tem[Humire et~al.(2022)]{Humire2022} Humire, P.~K., et al. 2022, \aap, 663, A33
\bibitem[Lee \& Seung(1999)]{Lee1999} Lee, D.~D., \& Seung, H.~S. 1999, \nat, 401, 788
\bibitem[Lee \& Seung(2001)]{Lee2001} Lee, D.~D., \& Seung, H.~S. 1999, Adv. Neural Inf. Proc. Syst., 13, 556
\bibitem[Mart\'{i}n et~al.(2021)]{Martin2021} Mart\'{i}n,~S., et al. 2021, \aap, 656, A46
\bibitem[Rekola et~al.(2005)]{Rekola2005} Rekola, R., Richer, M.~G., McCall, M.~L., Valtonen, M.~J., Kotilainen, J.~K., \& Flynn,~C. 2005, \mnras, 361, 330
\bibitem[Rico-Villas et~al.(2020)]{Rico2020} Rico-Villas,~F., Mart\'{i}n-Pintado,~J., Gonz\'{a}lez-Alfonso,~E., Mart\'{i}n,~S., \& Rivilla, V.~M. 2020, \mnras, 491, 4573
\bibitem[Tanaka et~al.(in preparation)]{Tanaka} Tanaka, et al. (in preparation)
\bibitem[Turner \& Ho(1985)]{Turner1985} Turner, J.~L., \& Ho, P. T.~P. 1985, \apj, 299, L77
\bibitem[Watanabe et~al.(2016)]{Watanabe2016} Watanabe,~Y., Sakai,~N., Sorai,~K., Ueda,~J., \& Yamamoto, S. 2016, \apj, 819, 144
\end{thebibliography}
\bibliographystyle{aasjournal}

\end{document}